\documentclass[prd,twocolumn,showkeys,floatfix,footinbib]{revtex4}
\usepackage{graphicx} 
\usepackage{amssymb} 
\usepackage{amsmath}
\usepackage{epsfig}
\newcommand{\be}{\begin{equation}} \newcommand{\ee}{\end{equation}}
\newcommand{\ba}{\begin{eqnarray}} \newcommand{\ea}{\end{eqnarray}}
\newcommand{\bea}{\begin{eqnarray}} \newcommand{\eea}{\end{eqnarray}}
\newcommand{\bean}{\begin{eqnarray*}} \newcommand{\eean}{\end{eqnarray*}}

\usepackage{slashed} 

\newcommand{\tr}[1]{\ensuremath{{\rm Tr}_{\rm D} \left[ #1 \right]}}
\newcommand{\trc}[1]{\ensuremath{{\rm Tr}_{\rm C} \left[ #1 \right]}}

\begin{document}

\title{No Generalized TMD-Factorization in the Hadro-Production of High Transverse Momentum Hadrons}  
\author{Ted C. Rogers}
\email{trogers@few.vu.nl}
\author{Piet J. Mulders}
\email{mulders@few.vu.nl}
\affiliation{Department of Physics and Astronomy,\\ Vrije Universiteit Amsterdam,\\ NL-1081 HV Amsterdam, The Netherlands}
\date{\today}

\begin{abstract}
It has by now been established that standard QCD factorization using  
transverse momentum dependent parton distribution functions fails in hadro-production 
of nearly back-to-back hadrons with high transverse momentum.  The 
essential problem is that gauge invariant transverse momentum dependent parton distribution 
functions cannot be defined with process-independent Wilson line operators, thus implying a breakdown  
of universality.  
This has led naturally to proposals that a correct approach 
is to instead use a type of ``generalized'' transverse momentum dependent factorization in which the basic factorized structure 
is assumed to remain valid, but with transverse momentum dependent 
parton distribution functions that contain non-standard, process dependent
Wilson line structures.  
In other words, to recover a factorization formula, it has become common to assume that it is sufficient to simply modify the Wilson lines
in the parton correlation functions for each separate hadron.
In this paper, we will illustrate by direct counter-example that this is not  
possible in a non-Abelian gauge theory.  
Since a proof of generalized transverse momentum dependent factorization should apply generally to any hard hadro-production process, a 
single counter-example suffices to show that a general proof does not exist.
Therefore, to make the counter-argument clear and explicit, we illustrate with a specific calculation for a 
double spin asymmetry in a spectator model with a non-Abelian gauge field.  
The observed breakdown of generalized transverse momentum dependent factorization challenges the notion 
that the role of parton transverse momentum in such processes can be described using 
separate correlation functions for each external hadron.
\end{abstract}

\maketitle

\section{Introduction}
\label{intro}
There has been much recent activity devoted to the study of parton 
transverse momentum in high energy hadronic collisions. 
Observables that are sensitive to parton transverse momentum can 
potentially provide new insight into the structure of 
hadrons. 
However, to interpret data it is important to understand the 
extent to which familiar perturbative QCD (pQCD) factorization approaches can be extended to 
situations that involve parton transverse momentum. 
In this paper, we specifically address recent efforts to apply the usual general framework of  
pQCD factorization (with some modifications) 
to describe parton transverse momentum in collisions between high energy 
hadrons with production of a pair of nearly back-to-back high transverse momentum hadrons 
or jets in the final state: 
\begin{equation} 
\label{eq:reaction}
H_1 + H_2 \to H_3 + H_4 + X. 
\end{equation} 
The distribution of transverse momentum inside the colliding hadrons 
can be probed, for example, by measuring the small imbalance in the distribution of transverse
momentum between the final state pair (see, e.g., ~\cite{Boer:2009nc,Lu:2008qu}).  Parton transverse momentum 
also plays a central role in the generation of spin asymmetries.

For a discussion of the relevant issues, it
will be important to first clearly define our terminology.
Very generally, a QCD factorization theorem~\cite{Collins:1989gx} is said to be valid if an observable 
(such as a cross section) can be written as a convolution product of 
factors that describe different regions of parton momentum.
Schematically, one expects for process~(\ref{eq:reaction}):
\begin{equation}
\label{eq:factorization}
d \sigma = \mathcal{H} \, \otimes \, \Phi_{H_1} \, \otimes \, \Phi_{H_2} \, \otimes \, \Delta_{H_3} \, \otimes \, \Delta_{H_4} + {\rm p.s.c.}
\end{equation}
Here, the symbol $\otimes$ denotes all relevant convolution integrals and traces, and 
a sum over different types of partons and subprocesses is understood.
The hard factor $\mathcal{H}$ describes the short range behavior in the hard collision between partons, 
while $\Phi_{H_1}$ and $\Phi_{H_2}$ are the parton distribution functions (PDFs) for 
initial state hadrons $H_1$ and $H_2$.  The $\Delta_{H_3}$ and $\Delta_{H_4}$ are fragmentation 
functions for final state hadrons $H_3$ and $H_4$.
In general, a soft factor may also be needed, though we will not write it explicitly.
The symbol ``p.s.c.'' indicates that power suppressed corrections are neglected.

A \emph{transverse momentum dependent (TMD)} factorization theorem
(also called unintegrated factorization or $k_T$-factorization) is said to be
valid if the role of parton transverse momentum can be taken into account 
by using PDFs and FFs in Eq.~(\ref{eq:factorization}) which depend explicitly on parton 
transverse momentum.  In a TMD-factorization formula,  
one therefore refers to TMD PDFs and TMD FFs (or ``unintegrated'' PDFs and FFs).
 
TMD-factorization should be contrasted with the more common \emph{collinear} 
factorization theorems, applicable to cases where observables are not sensitive to intrinsic transverse parton momentum. 
In the collinear factorization theorems, transverse momentum is 
integrated over inside the definitions of the standard PDFs and FFs.
These are the standard ``integrated''
PDFs and FFs which are expressible as well-defined, process-independent operator matrix elements~\cite{Collins:1981uw}.  
The process-independence means that the integrated PDFs and FFs can be 
parameterized by experimental data and then later 
reused in calculations to make first-principle predictions for future experiments.  
Thus, this universality property of the standard integrated PDFs and FFs lends great 
predictive power to the standard pQCD approaches, and is a basic component of the standard collinear factorization theorems. 
  
It is natural to hope that an analogous universality property applies to the TMD PDFs and TMD FFs in a TMD-factorization formula.  
In that case, we would say that \emph{standard} TMD-factorization is valid, where we will explicitly 
use the word ``standard'' to refer to the universality condition. 
If a standard TMD-factorization formula were valid for the process in~(\ref{eq:reaction}), 
then it would be possible to calculate cross sections using the same TMD PDFs and FFs that are parameterized in other processes like 
deep inelastic scattering (DIS) or the Drell-Yan (DY) process.
However, as we will discuss in more detail, TMD-factorization is not generally valid
for the process in Eq.~(\ref{eq:reaction}).

The essential complication arises from longitudinally polarized gluons that couple 
soft and collinear subgraphs to the hard part and which, at first sight, 
appear to break topological factorization graph-by-graph at leading 
power.
Dealing with these ``extra'' gluons is one of the main issues that must be dealt with in all factorization proofs in pQCD.  
In DIS, for example, it is found that after summing over all
graphs (and applying appropriate approximations), a Ward identity argument allows the extra gluon contributions 
to be factored into contributions which correspond to Wilson lines (also called gauge links) in the definitions 
of the PDFs or FFs~\cite{Collins:1989gx}.  
However, in order to justify the approximations that allow an application of the
Ward identity, it is necessary to first perform certain contour deformations on 
the gluon momentum~\cite{Collins:1985ue}.  For DIS, the contour deformations must be consistent with having 
extra gluon attachments between hadron spectators and a final state struck quark.
Similar steps apply to DY, but there the contour deformations should correspond to 
extra attachments between hadron spectators and an initial state quark.
(See also Ref.~\cite{Collins:2004nx}.)
The result is that the Wilson line for the quark TMD PDF in DIS is future pointing while  
the Wilson line for a quark TMD PDF in DY is past pointing~\cite{Collins:2002kn}.  The difference 
in Wilson line direction has been shown to result in a sign flip for the Sivers function (a particular type of TMD PDF)
in DY as compared to DIS~\cite{Collins:2002kn}.  Strictly speaking, this could be regarded as a breakdown of standard 
TMD-factorization because the Sivers function is not truly universal in DIS and DY. 
However, since a sign flip is easily accounted for, it is 
more appropriate to say that the Sivers function in DIS and DY possesses a type of ``modified'' universality.
The relationship between correlation functions in DIS and DY is further discussed in Ref.~\cite{Boer:2003cm}.

The problems that occur with TMD-factorization for Eq.~(\ref{eq:reaction}) are much more complicated because there one must deal with
extra gluons that connect spectators to both initial and final state partons.  
The result is that the contour deformations necessary for a factorization proof prevent 
a direct application of the usual Ward identity arguments.   
This problem was observed by Bomhof, Mulders and Pijlman~\cite{Bomhof:2004aw}
who found that the Wilson lines needed for gauge-invariant TMD PDFs and FFs in hadro-production of hadrons
are not generally process-independent.  
Although the extra gluon attachments eikonalize, the resulting sums of eikonal factors 
do not correspond to the simple future or past pointing Wilson lines that are found in DIS and DY.
In~\cite{Bomhof:2004aw} it was shown at the level of a single extra gluon that the 
TMD PDFs and FFs (assuming consistent definitions exist) are non-universal because they require, at a minimum, process dependent Wilson lines. 
Hence, there is a violation of the universality property necessary for standard TMD-factorization.

Similar problems are encountered in proofs of collinear factorization 
for hadron-hadron scattering, but there one 
is saved by cancellations between graphs that occur \emph{after} integration over transverse momentum (see, e.g., Ref.~\cite{Collins:1985ue} 
and also more recent work in Ref.~\cite{Aybat:2008ct}).
Since these cancellations are not point-by-point in transverse momentum, they do not generally apply to TMD-factorization.
  
To address the role of non-universal Wilson lines, the concept of a ``generalized'' TMD-factorization formula was 
later developed~\cite{Bacchetta:2005rm,Bomhof:2006dp}.
In this approach, it is assumed that the only deviation from 
standard TMD-factorization is that the TMD parton 
correlation functions (PDFs and FFs) must contain process-dependent Wilson lines.
Schematically, instead of Eq.~(\ref{eq:factorization}) one must assume
a more complicated expression of the form
\begin{multline}
\label{eq:genfactor}
d \sigma = \\ \sum_{j,c} \mathcal{H}_{j,c} \otimes \,
\Phi_{H_{1},j}^{[WL_{1}^{c,j}]} \, \otimes \, \Phi_{H_{2},j}^{[WL_{2}^{c,j}]} 
\, \otimes \, \Delta_{H_{3},j}^{[WL_{3}^{c,j}]} \, \otimes \, \Delta_{H_{4},j}^{[WL_{4}^{c,j}]} \\ + {\rm p.s.c.} 
\end{multline}
Here, $\mathcal{H}_{c,j}$ is the hard part for subprocess $j$ and color routing $c$.
For each subprocess and routing of color through the hard part, there is in general a different set of 
TMD PDFs and TMD FFs corresponding to the different 
Wilson line structures that they contain. In the above notation $\Phi_{H_{1},j}^{[WL_{1}^{c,j}]}$ is 
a gauge invariant TMD PDF for hadron $H_1$ with 
a Wilson line $[WL_{1}^{c,j}]$ corresponding to subprocess $j$ and color routing $c$.
Analogous notation is used for the other process-dependent
correlation functions.  
(There are also possible soft factors not shown explicitly in Eq.~(\ref{eq:genfactor})).  
Each term in Eq.~(\ref{eq:genfactor}) has the same basic factorized structure 
as in Eq.~(\ref{eq:factorization}), involving distinct (though process-dependent) 
TMD PDFs and FFs for all external hadrons.
The TMD PDFs and FFs have the usual structure of a pair of field operators and a Wilson line with an  
expectation value corresponding to a specific external hadron state.
The only difference from the standard case is that they are equipped with non-standard and potentially complex Wilson line structures.
In particular, Eq.~(\ref{eq:genfactor}) contains no matrix element of the 
form $\langle H_1 H_2 | \cdots | H_1 H_2 \rangle$. 
So, by \emph{generalized} TMD-factorization we mean that a TMD-factorization formula is recovered 
simply by replacing the Wilson lines in the definitions of the correlation functions by non-standard ones, which 
may be different for each hard subprocess and for each way of routing color through the hard part.
The TMD PDFs needed in a generalized TMD-factorization formula for Eq.~(\ref{eq:reaction}) could be totally different from the 
ones parameterized in, e.g. DIS and DY.  

A conjectured TMD-factorization of the form of 
Eq.~(\ref{eq:genfactor}) is a basic assumption in a number of 
recent studies~\cite{Bacchetta:2005rm,Bomhof:2006dp,Bomhof:2007su,Bacchetta:2007sz,Bomhof:2007xt,Boer:2007nd}.  
The minimal Wilson line structures needed for Eq.~(\ref{eq:genfactor})
can be determined by considering a single extra gluon at a time, radiated from each of the external hadrons and 
attaching everywhere in the hard subprocess.  The resulting process-dependent gauge invariant correlation functions 
have been tabulated in Refs.~\cite{Bacchetta:2005rm,Bomhof:2006dp}.
These correlation functions have also been used to 
calculate physical observables such as weighted spin asymmetries~\cite{Bomhof:2007su,Bacchetta:2007sz}. 

Collins and Qiu~\cite{collins.qiu} verified explicitly 
that \emph{standard} TMD-factorization fails in a sample calculation of 
a single spin asymmetry (SSA).  That is, they showed in an explicit calculation that the process-dependence of the
Wilson line structures observed in Ref.~\cite{Bomhof:2004aw} indeed corresponds to non-universality 
for the TMD PDFs.  For their calculation, they used a 
model Abelian theory and calculated the effect of a single extra gluon.   
An explicit illustration of the violation of standard TMD-factorization was also given  
for unpolarized scattering in Ref.~\cite{Collins:2007jp}, again using a model Abelian gauge theory.
The two-gluon example for unpolarized scattering is important as it 
directly illustrates that standard TMD-factorization cannot generally be recovered 
by rescaling the hard part with a constant color factor. (Compare this with the procedure of Refs.~\cite{Qiu:2007ar}.)
In Ref.~\cite{Vogelsang:2007jk}, it was shown explicitly that the observed breakdown of 
standard TMD-factorization described in Refs.~\cite{collins.qiu,Collins:2007jp} is 
consistent with the generalized TMD-factorization 
proposed in Refs.~\cite{Bacchetta:2005rm,Bomhof:2006dp,Bomhof:2007xt}, again within the Abelian theory.

However, all the cases studied so far have only considered graphs with extra gluons 
radiated from \emph{one} of the hadrons at a time.  
What is missing is a treatment of non-Abelian gluons radiated from different hadrons simultaneously.
If a generalized TMD-factorization approach is possible, then extra gluons radiated from all 
hadrons simultaneously must be shown to eikonalize and factorize after a sum over graphs.   
Given the complex color structures that arise in a non-Abelian gauge theory, 
it is unclear that such a procedure is possible in real QCD.

The purpose of this paper is to show explicitly that even generalized 
TMD-factorization breaks down in a non-Abelian gauge theory at 
the level of two extra gluons.   
In other words, the violation of standard TMD-factorization, already found in previous 
work, cannot be dealt with simply by replacing the Wilson lines in the 
standard correlation functions by more complicated ones and summing over different subprocesses and 
color structures as in Eq.~(\ref{eq:genfactor}). 

As seen in Ref.~\cite{collins.qiu},
the basic reasons for a breakdown of \emph{standard} TMD-factorization are illustrated 
most directly in a calculation of an SSA with a single extra gluon. 
We will find analogously that the breakdown of \emph{generalized} TMD-factorization in 
a non-Abelian gauge theory is most easily illustrated in a 
calculation of a double Sivers effect in a double transverse spin asymmetry (DSA).  As in 
Refs.~\cite{collins.qiu,Collins:2007jp} we will use a model field theory to describe 
the quarks, spectators and hadrons.
A proper counterexample to generalized TMD-factorization must verify that terms which 
violate generalized TMD-factorization graph-by-graph do not cancel in a sum over graphs.
This is most easily done in a simple spectator model that restricts the number of relevant Feynman graphs.  

In Sect.~\ref{sec:setup} we discuss the particular model and describe 
the procedure for deriving a violation of generalized TMD-factorization.  In Sect.~\ref{sec:onegluon} we 
review the steps for factorization with one extra gluon.  We explicitly review the breakdown 
of standard TMD-factorization for two extra gluons from one hadron in Sect.~\ref{sec:twogluons}.
In Sect.~\ref{sec:genfact} we discuss the generalized TMD-factorization formula that is required to recover a
factorized structure.  In Sect.~\ref{sec:breakdown} we demonstrate that the generalized TMD-factorization formula
is inconsistent with having extra gluons radiated from both hadrons simultaneously.  We end with concluding 
remarks in Sect.~\ref{sec:conc}.

\section{Setup}
\label{sec:setup}

\begin{figure*}
\centering
\includegraphics[scale=.5]{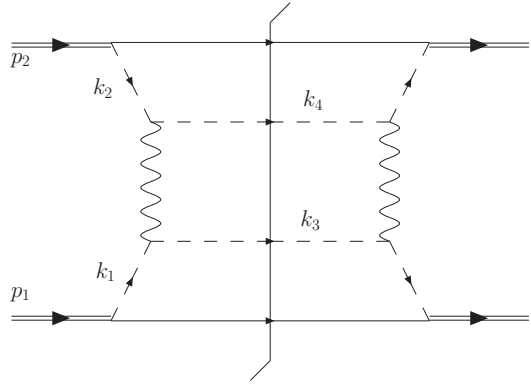}
\caption{Basic graph contributing to hadro-production with no extra gluons.}
\label{fig:basic}
\end{figure*}
A simple model field theory provides a direct illustration of 
why factorization fails in a gauge theory, while avoiding the complications 
of dealing with a large number of Feynman graphs.   
We will continue to use the model field theory of 
Refs.~\cite{collins.qiu,Collins:2007jp,Brodsky:2002cx}, though with a few important differences.
The hadrons continue to correspond to different flavors.
The ``quarks'' continue to correspond to scalar fields $\phi_f$, while 
the ``hadron'' fields $H_f$ and the spectator ``diquarks'' $\psi_f$ 
are Dirac spinors.  The subscript $f = 1,2$ labels flavor.   
The main difference from Refs.~\cite{collins.qiu,Collins:2007jp,Brodsky:2002cx} which we will introduce is that the gauge field will be 
the massless ${\rm SU}(N_c)$ non-Abelian gauge field.  (In QCD $N_c = 3$.)  
By contrast, Refs.~\cite{collins.qiu,Collins:2007jp} used a massive 
Abelian gauge field that coupled with different 
charges, $g_1$ and $g_2$, to the quarks and diquarks in hadrons $H_1$ and $H_2$.
In this paper, the non-Abelian gauge 
field couples with equal strength to the quarks in each hadron, $g_1 = g_2 = g$.

We also will introduce one more field: The hard subprocess will be described by an exchange 
of a hypothetical new massive color-neutral ${\rm U}(1)$ gauge boson which we will call $X$.  
It couples with strengths $\lambda^\prime_{1(2)}$ to quarks of flavor $1(2)$.  
Hence, the hard subgraph has a trivial color structure.  This will allow us to study the role of 
extra gluons in a non-Abelian gauge theory, while limiting the number of Feynman graphs that need to be explicitly considered, 
and eliminating color flow directly through the hard part.
It is important to emphasize that the general conclusions of this paper do not depend on this specific model for the hard subprocess.
What is important is that color is carried by both the initial and final state partons.  
A proof of generalized TMD-factorization for such processes should be general and apply to any hard hadro-production 
process with observed hadrons in both the initial and final states.  If it fails for a specific example, 
then no general proof exists.
In cases where color is exchanged in the hard process, the Wilson line structures for the TMD PDFs in Eq.~(\ref{eq:genfactor}) 
are determined after first expanding 
the color matrices of the hard subprocess into different routings following the procedure described in Ref.~\cite{Bomhof:2006dp}.
The steps in this paper then apply to each term in a series of color routings, 
and similar contradictions with generalized TMD-factorization can be found.
Our choice of hard subprocess is to make the illustration of why generalized TMD-factorization fails as clear as possible.

The basic graph contributing to hadro-production of high-$p_t$ hadrons, 
with no extra gluon attachments, is shown in Fig.~\ref{fig:basic}.
Because of the trivial color structure of the hard part, and the simplicity of our model of quarks and spectators,
there is only one subprocess and one color flow.  So a generalized TMD-factorization 
formula corresponding to Eq.~(\ref{eq:genfactor}) can only involve
one term.

We work in the center-of-mass frame where the incoming 
hadron $H_1$ is initially moving with large rapidity in the 
forward plus direction with no transverse momentum, while $H_2$ is initially moving 
with large rapidity in the minus direction with no transverse momentum.  
Each hadron splits into an active ``quark'' which enters the hard 
subgraph, and a spectator ``diquark'' which enters the 
final state.  The quarks interact in the hard part by exchanging a hard 
colorless vector boson $X$, with momentum $q$ where $|q^2|$ is large, $|q^2| \gg M_X^2$.
Within the model, the final state high transverse momentum hadrons or jets 
are represented simply by on-shell final state quarks.

The polarization dependent differential cross section at zeroth order in $g$ is,
\begin{widetext}
\begin{multline}
\label{eq:basiceq}
E_3 E_4 \frac{d \sigma}{d^3 {\bf k_3} \, d^3 {\bf k_4} } = \frac{\lambda_1^2 \lambda_2^2 {\lambda^\prime_1}^2 {\lambda^\prime_2}^2 N_c^2}{8 s (2 \pi)^6 }
\int \frac{d^4 k_1}{(2 \pi)^4} \frac{d^4 k_2}{(2 \pi)^4} \, (2 \pi)^4 \delta^4 (k_1 + k_2 - k_3 - k_4) 
\left\{ \frac{ (k_1 + k_3) \cdot (k_2 + k_4) }{(k_1 - k_3)^2 - M_X^2} \right\}^2  \times \\ \times  
\frac{\frac{1}{2} \tr{(\slashed{p}_1 + m_{H_1})  (1 + \gamma_5 \slashed{s}_1 ) (\slashed{p}_1 - \slashed{k}_1 + m_{\psi_1}) } }{(k_1^2 - m_{q_1}^2)^2}
\frac{\frac{1}{2} \tr{(\slashed{p}_2 + m_{H_2})  (1 + \gamma_5 \slashed{s}_2 ) (\slashed{p}_2 - \slashed{k}_2 + m_{\psi_2}) } }{(k_2^2 - m_{q_2}^2)^2}
\times \\ \times
(2 \pi)^2 \delta ((p_1 - k_1)^2 - m_{\psi_1}^2 ) \delta ((p_2 - k_2)^2 - m_{\psi_2}^2 ). 
\end{multline}
The factor of $N_c^2$ comes from tracing over the unit color triplet matrices in the upper and lower color loops in Fig.~\ref{fig:basic}.
We will use the approximation that the struck partons are collinear to the directions of their parent hadrons.  
So, $k_1^+ \sim p_1^+$ and $k_2^- \sim p_2^-$.  The minus component of $k_1$ and the plus component 
of $k_2$ are small (order $\Lambda^2/p_1^+$) and can be neglected inside the hard part.
We also define $x_1 = k_1^+ / p_1^+$ and $x_2 = k_2^- / p_2^-$, and $s$ is the usual Mandelstam variable for center-of-mass energy squared.
The transverse spin vectors $s_{1(2)}$ label the transverse spin for hadrons $H_{1(2)}$ and are normalized such that the extreme values 
are $s_{1(2)}^2 = 1$.
The subscript $D$ on $\tr{\cdots}$ indicates a trace over Dirac indices.

\begin{figure*}
\centering
  \begin{tabular}{c@{\hspace*{5mm}}c}
    \includegraphics[scale=0.5]{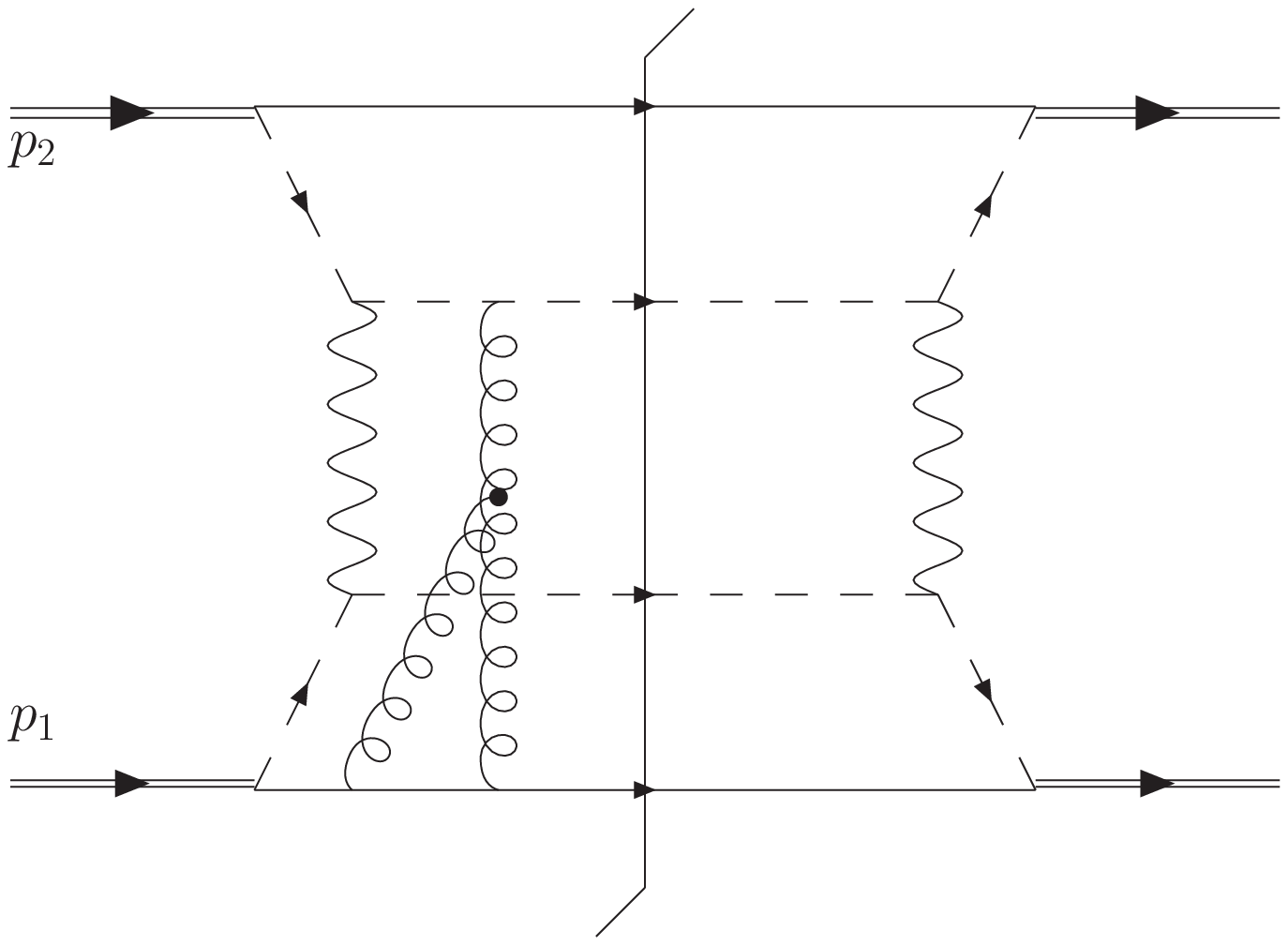}
    &
    \includegraphics[scale=0.5]{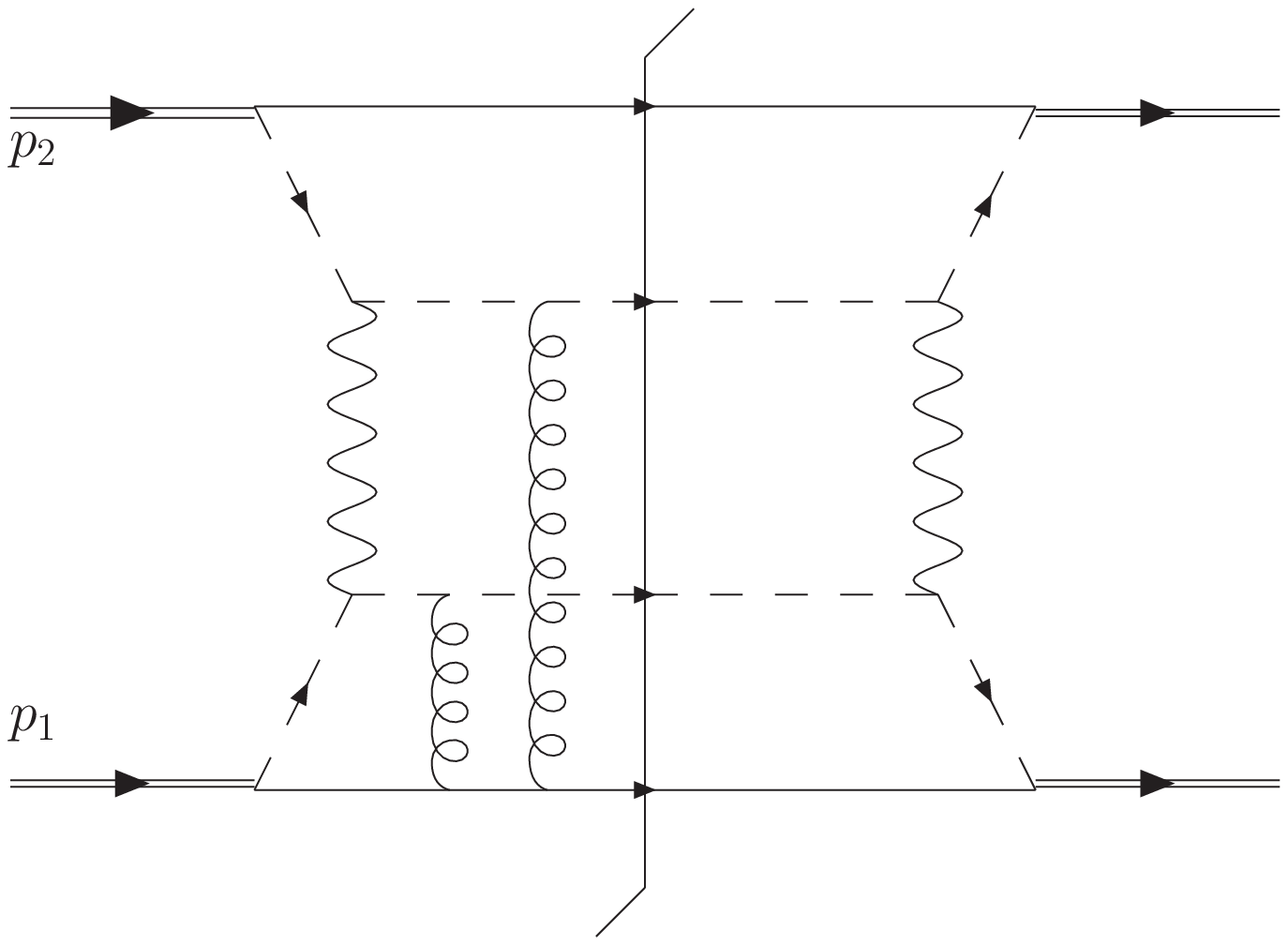}
  \\
  (a) & (b)
  \end{tabular}
\caption{Typical cases of graphs that vanish when extra gluons are considered because of the trivial color 
factor, $\trc{t^a} = 0$.}
\label{fig:zero}
\end{figure*}

After evaluating the $\delta$-functions, this cross section easily factorizes:
\begin{multline}
\label{eq:zerothcross}
E_3 E_4 \frac{d \sigma}{d^3 {\bf k_3} \, d^3 {\bf k_4} } = 
\frac{1}{2 s}  
\int \frac{d^2 {\bf k}_{1T}}{(2 \pi)^2} \mathcal{H}(k_1,k_3,k_4) \times \\ \times
\left\{ \frac{N_c \lambda_1^2 x_1 (1 - x_1)}{16 \pi^3} \frac{ \frac{1}{2} \tr{ (\slashed{p}_1 + m_{H_1})  
(1 + \gamma_5 \slashed{s}_1 ) (\slashed{p}_1 - \slashed{k}_1 + m_{\psi_1} ) } }{\left[ k_{1T}^2 - x_1 (1 - x_1) m_{H_1}^2 
+ (1 - x_1) m_{q_1}^2 + x_1 m_{\psi_2}^2 \right]^2} \right\} \times \\ \times
\left\{ \frac{N_c \lambda_2^2 x_2 (1 - x_2)}{16 \pi^3} \frac{\frac{1}{2} \tr{(\slashed{p}_2 + m_{H_2})  
(1 + \gamma_5 \slashed{s}_2 ) (\slashed{p}_2 - \slashed{k}_2 + m_{\psi_2}) } }{\left[k_{2T}^2 - x_2 (1 - x_2) m_{H_2}^2 
+ (1 - x_2) m_{q_2}^2 + x_2 m_{\psi_2}^2 \right]^2} \right\}, 
\end{multline}
where the hard part is,
\begin{equation}
\label{eq:hard}
\mathcal{H}(k_1,k_3,k_4) = \frac{{\lambda^\prime_1}^2 
{\lambda^\prime_2}^2}{2 x_1 x_2 s} \left\{ \frac{ (k_1 + k_3) \cdot (k_3 + 2 k_4 - k_1) }{(k_1 - k_3)^2 - M_X^2} \right\}^2. 
\end{equation}
The remaining factors in braces in Eq.~(\ref{eq:zerothcross}) 
are what is expected from the zeroth order expansion of the standard operator definition 
of a TMD PDF.  For hadron $H_1$, for example, the zeroth order TMD PDF is
\begin{equation}
\left[ \Phi_{H_1}(x_1,k_{1T}) \right]^{\mathcal{O}(1)} =
\frac{N_c \lambda_1^2 x_1 (1 - x_1)}{16 \pi^3} \times \frac{ \frac{1}{2} \tr{ (\slashed{p}_1 + m_{H_1})  
(1 + \gamma_5 \slashed{s}_1 ) (\slashed{p}_1 - \slashed{k}_1 + m_{\psi_1} ) } }{\left[ k_{1T}^2 - x_1 (1 - x_1) m_{H_1}^2 
+ (1 - x_1) m_{q_1}^2 + x_1 m_{\psi_2}^2 \right]^2} 
\end{equation}
\end{widetext}
and similarly for hadron $H_2$.  Note that ${\bf k}_{2T}$ is not kinematically independent
of ${\bf k}_{1T}$: ${\bf k}_{2T} = {\bf k}_{3T} + {\bf k}_{4T} - {\bf k}_{1T}$.  Also, $x_1 \approx (k_3^+ + k_4^+) / p_1^+$ and 
$x_2 \approx (k_3^- + k_4^-) / p_2^-$.  To keep the expression simple, we have written Eq.~(\ref{eq:hard}) in terms of exact 
parton momenta, $k_1$,$k_2$,$k_3$,$k_4$, though we remark that in the correct final formula these should be replaced by the approximate 
on-shell parton momenta appropriate for a hard subgraph~\footnote{Because in Eq.~(\ref{eq:basiceq}) there is an integration over 
all $k_{1T}$, one generally must be careful in treating the region where the $t$ goes to zero in the hard part.  For this paper, we are 
specifically interested in the collinear region where $k_{1T}$ is small, so we will assume that any contribution 
from $t$ close to zero is removed by a cutoff on large $k_{1T}$.}.

When treating higher order graphs, we will always work in Feynman gauge, where the contour 
deformations needed in a derivation of factorization are most straightforward.
For factorization to work, it must be possible to identify any uncanceled collinear or soft singularities  
as contributions to non-perturbative correlation functions.
Some of these singularities correspond to Wilson line contributions, and are therefore essential for maintaining 
gauge invariance.

Before continuing we should mention that
there are a number of general complications involved in deriving TMD-factorization that will not be addressed here because
they are not directly related to the main reasons that generalized TMD-factorization fails.  
The most naive definitions of TMD correlation functions include extra divergences that need to be removed by appropriate redefinitions 
(see Ref.~\cite{Collins:2008ht} and references therein for an overview of issues related to the precise definition of a TMD PDF).
One particular complication is that the Wilson lines in these definitions cannot be exactly 
light-like without containing extra ``light-cone'' divergences which 
correspond to partons moving with large rapidity opposite to the direction of their parent hadrons.  
Proposed solutions involve either tilting the Wilson line away from the exactly light-like direction~\cite{Collins:1981uw}, or 
dividing out by extra gauge invariant factors~\cite{Collins:2004nx,hautmann,Collins:2000gd,Collins:2003fm}.
Furthermore, an exactly correct factorization formula requires a soft factor to account for gluons with all components small.
The other parton correlation functions tend to overlap in the soft region, and a fully correct definition of a TMD PDF requires extra
factors to remove the overlap.

There has been much significant recent work devoted to finding a fully consistent definition of the TMD PDF.
However, our main concern in this paper is only with the 
general color structure of the Wilson lines inside the matrix elements for the external hadrons.
Therefore, for our purposes it will be sufficient to 
continue to treat the direction of the main Wilson lines in the TMD PDFs as being light-like, and we will not
address the role of a soft factor or the overlap of regions.
Finally, we will restrict consideration to the limit of very large relative transverse momentum (large-$p_t$) where one expects standard pQCD methods to 
be most appropriate, and we do not consider the possibility of recovering a type of 
TMD-factorization appropriate in the small-$x$ limit~\cite{Chang:2009bk}.

Having discussed the basic graph in Fig.~\ref{fig:basic}, the next step is to 
consider graphs dressed with extra gluons.
The graphs which can contribute to the Wilson line insertions are those in 
which extra, nearly on-shell gluons connect different subgraphs --- for example, 
graphs with gluons connecting the $p_1$-collinear lines to the outgoing 
struck quark lines or to the $p_2$-collinear lines.
Normally, one expects the sum of such graphs to contribute 
to the Wilson line in the TMD PDF for $H_1$, after application of a Ward identity.
The primary issue is that the approximations that normally allow Ward identities to be applied are only valid 
after certain contour deformations on the extra momentum integrals.
Namely, to apply a Ward identity, it must be possible to approximate
an extra gluon by a longitudinally polarized one with a large component of longitudinal momentum.
Then the extra gluon momentum can be contracted with the hard scattering matrix element and a Ward identity 
argument can be applied directly.
However, when virtual gluons attach to a spectator line, they give contributions from 
the ``Glauber'' region, meaning that if $l$ is the momentum 
of an extra gluon, then $|l^+|$ and $|l^-|$ are both much smaller than $|{\bf l}_{T}|$.  
In the Glauber region, the approximations needed for the Ward
identity are not valid.  For factorization to work, it must be possible to first  
deform the contour out of the Glauber region (see, for example, Refs.~\cite{Collins:2004nx}).  Alternatively, factorization could be recovered
if there is a cancellation between graphs, as in the standard proofs of integrated (collinear) factorization.
The basic problem with TMD-factorization found in previous work~\cite{Bacchetta:2005rm,Bomhof:2006dp,Bomhof:2007xt,collins.qiu,Collins:2007jp} 
is that the necessary contour deformations
needed to treat the Glauber region are inconsistent with a direct application of the standard Ward identity arguments.
Namely, they are in different directions for different graphs depending on whether the interaction is in the initial 
or final state.

So, we will only consider graphs that can yield contributions from the Glauber region.
Since real gluons can never be in the Glauber region, we will only consider graphs with virtual gluons.
Also, as long as no restrictions are placed on the target remnant momenta, graphs with spectator-spectator 
interactions 
cancel~\cite{Collins:1985ue} in the integration over final states~\footnote{In a general proof 
of factorization, one may 
need to worry about cancellations in the ``supersoft'' region, $k_{1T} << \Lambda_{QCD}$, since we are 
now using a massless theory, in contrast to Ref.~\cite{collins.qiu}.  In this paper, we are concerned mainly 
with gluons with $k_{1T} \sim \Lambda_{QCD}$ so these issues do not affect our argument.}. 
Similar cancellations occur between different cuts of the same graph for active-spectator 
interactions after parton transverse momentum is integrated over, and are needed in the standard 
proofs of collinear factorization~\cite{Collins:1985ue}.  A counter-proof of TMD-factorizaton therefore needs to show that such 
cancellations generally fail when transverse parton momentum is explicitly taken into account.  A specific example of such 
a non-cancellation was given in Ref.~\cite{Collins:2007jp} and will be reviewed in Sect.~\ref{sec:twogluons}.  
In graphs with attachments between active quarks, there are not enough Dirac $\gamma$-matrices
to give spin dependence to the TMD PDFs.  Such graphs will therefore not affect our discussion of single and double spin 
asymmetries at lowest non-vanishing order. 
Furthermore, graphs with a scalar-scalar-gluon-gluon vertex do not give 
leading power contributions to eikonal factors.

We remark that, because the TMD factorization breaking effects are due to the Glauber region where all components of 
gluon momentum are small, the interactions responsible for breaking TMD factorization are associated with large distance scales. 

In our specific model, a large number of graphs vanish simply because of the highly simple color structure involved.
Examples are shown in Fig.~\ref{fig:zero}.  They vanish because their color factors include a trace around a
color loop of a single ${\rm SU}(N_c)$ generator, $\trc{t^a} = 0$.  (The $C$ on the $\trc{\cdots}$ denotes a trace over 
triplet color indices.)

\begin{figure*}
\centering
  \begin{tabular}{c@{\hspace*{5mm}}c}
    \includegraphics[scale=0.5]{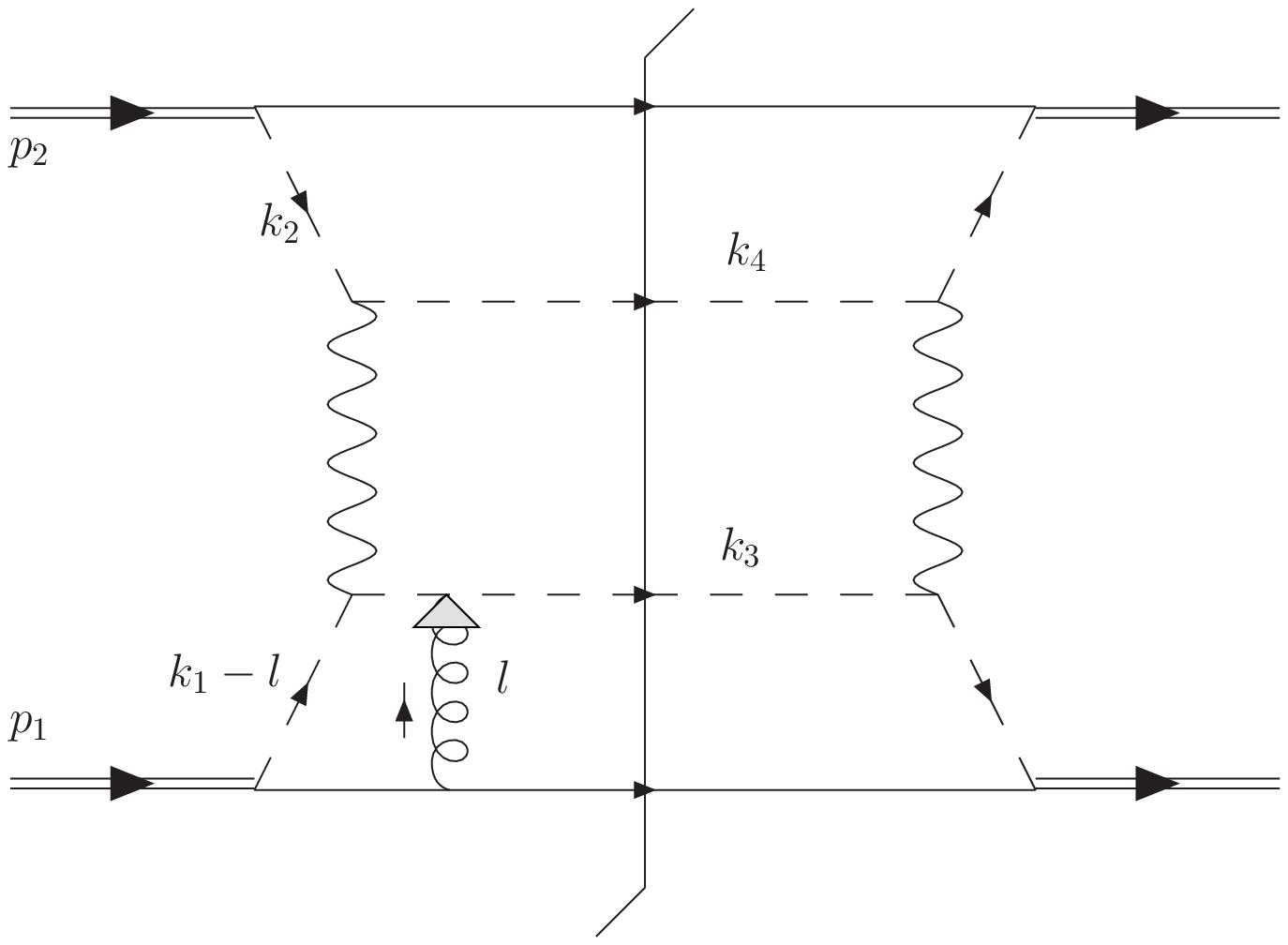}
    &
    \includegraphics[scale=0.5]{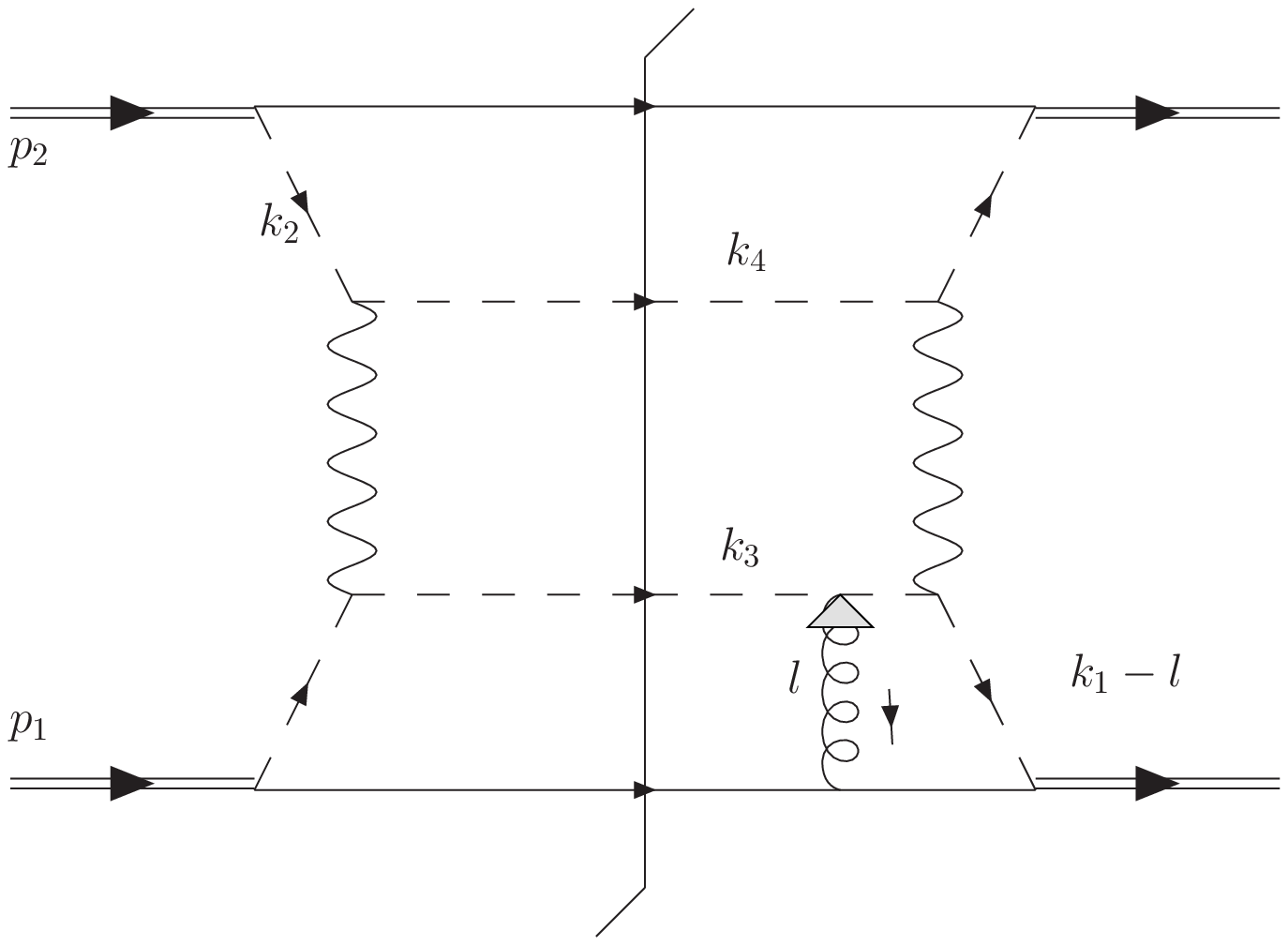}
  \\
  (a) & (b)
  \\[3mm]
    \includegraphics[scale=0.5]{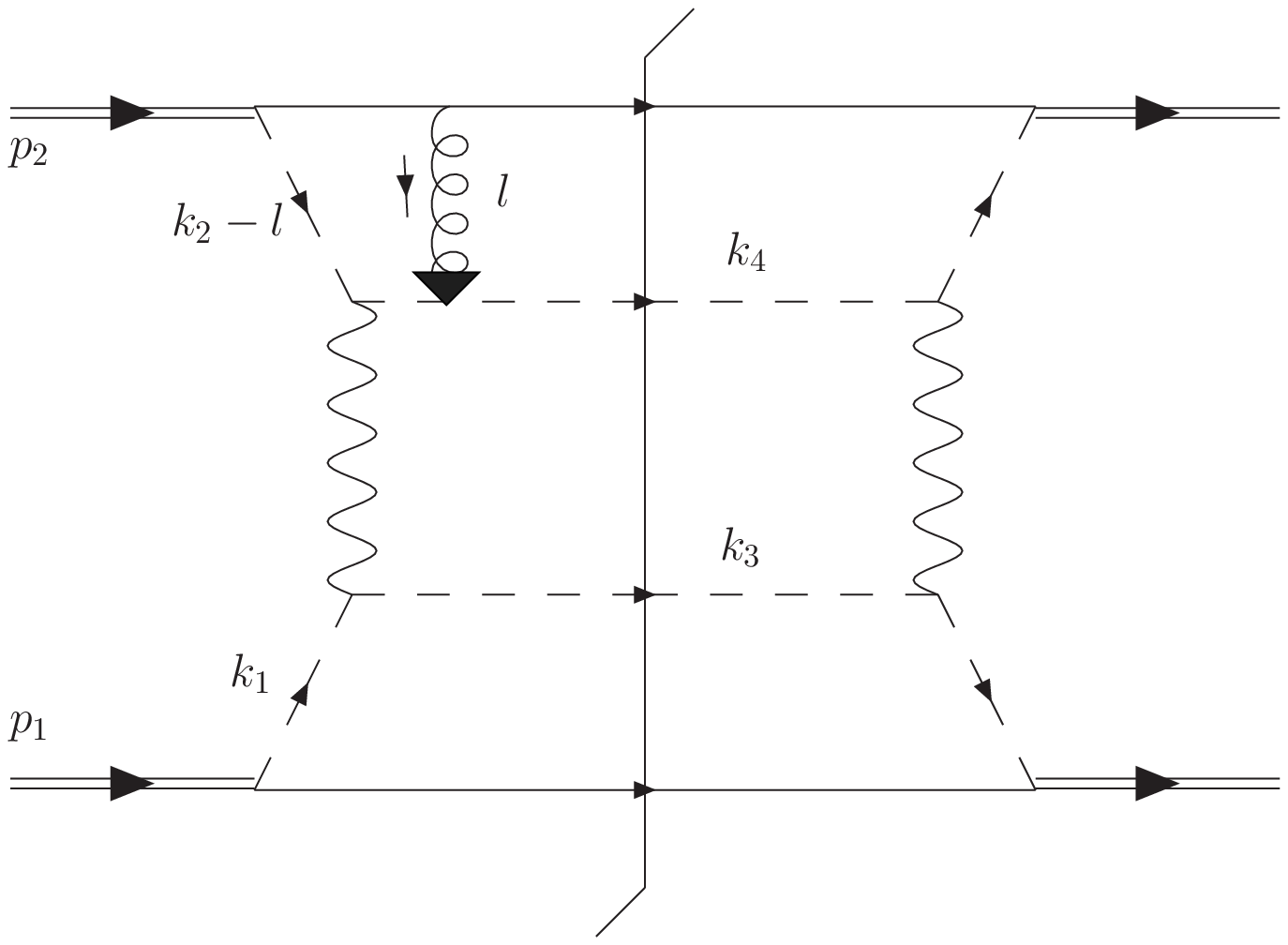}
    &
    \includegraphics[scale=0.5]{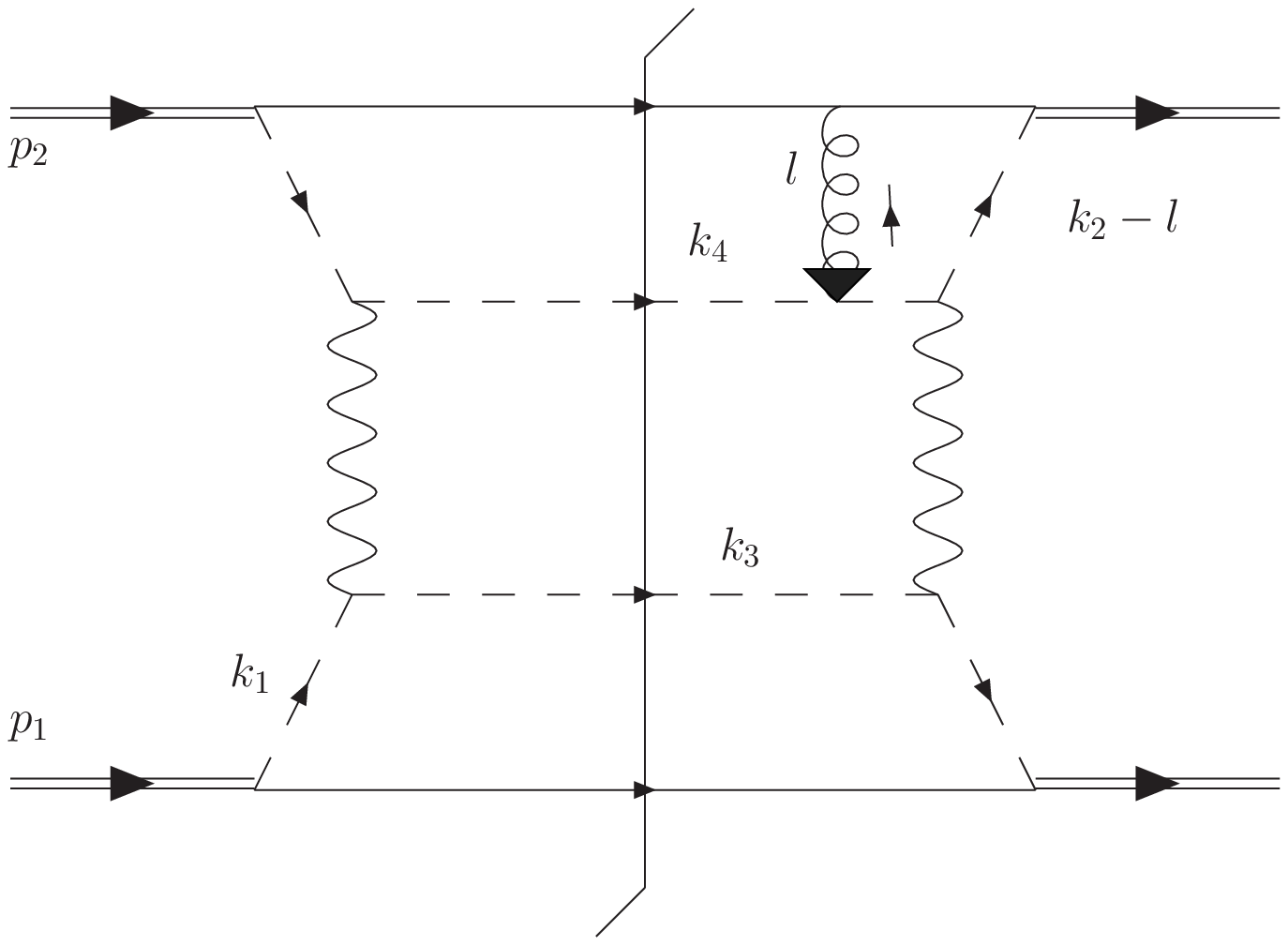}
   \\
   (c) & (d)
  \end{tabular}
\caption{Extra gluon attachments that are consistent with the standard Wilson line structure.  The difference in 
shading on the arrow at the eikonal attachments for (a,b) and (c,d) is to emphasize that the gluons are from different hadrons.
They correspond to $p_1$ in (a,b) and to $p_2$ in (c,d).}
\label{fig:onegluestand}
\end{figure*}

\begin{figure*}
\centering
  \begin{tabular}{c@{\hspace*{5mm}}c}
    \includegraphics[scale=0.5]{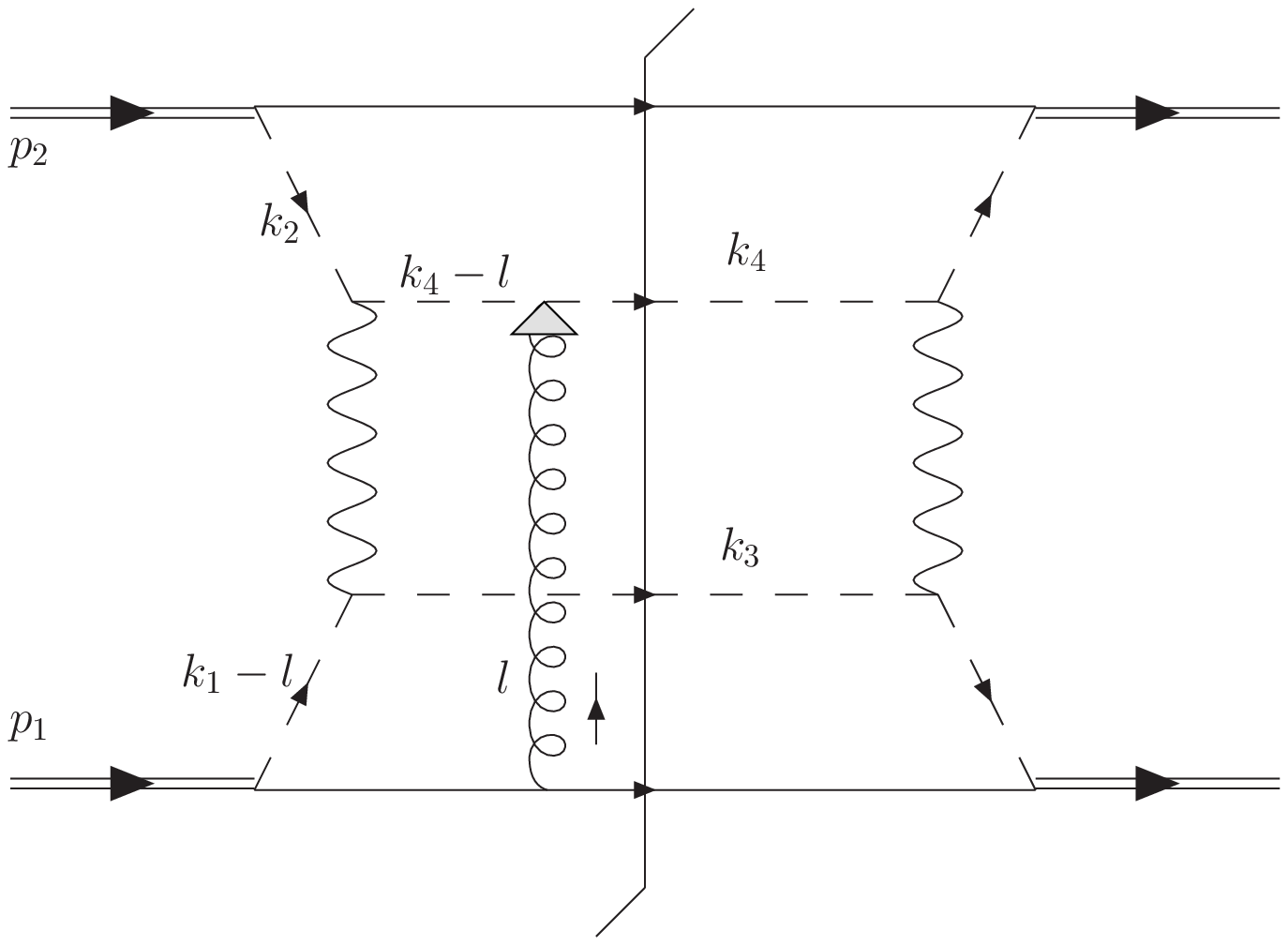}
    &
    \includegraphics[scale=0.5]{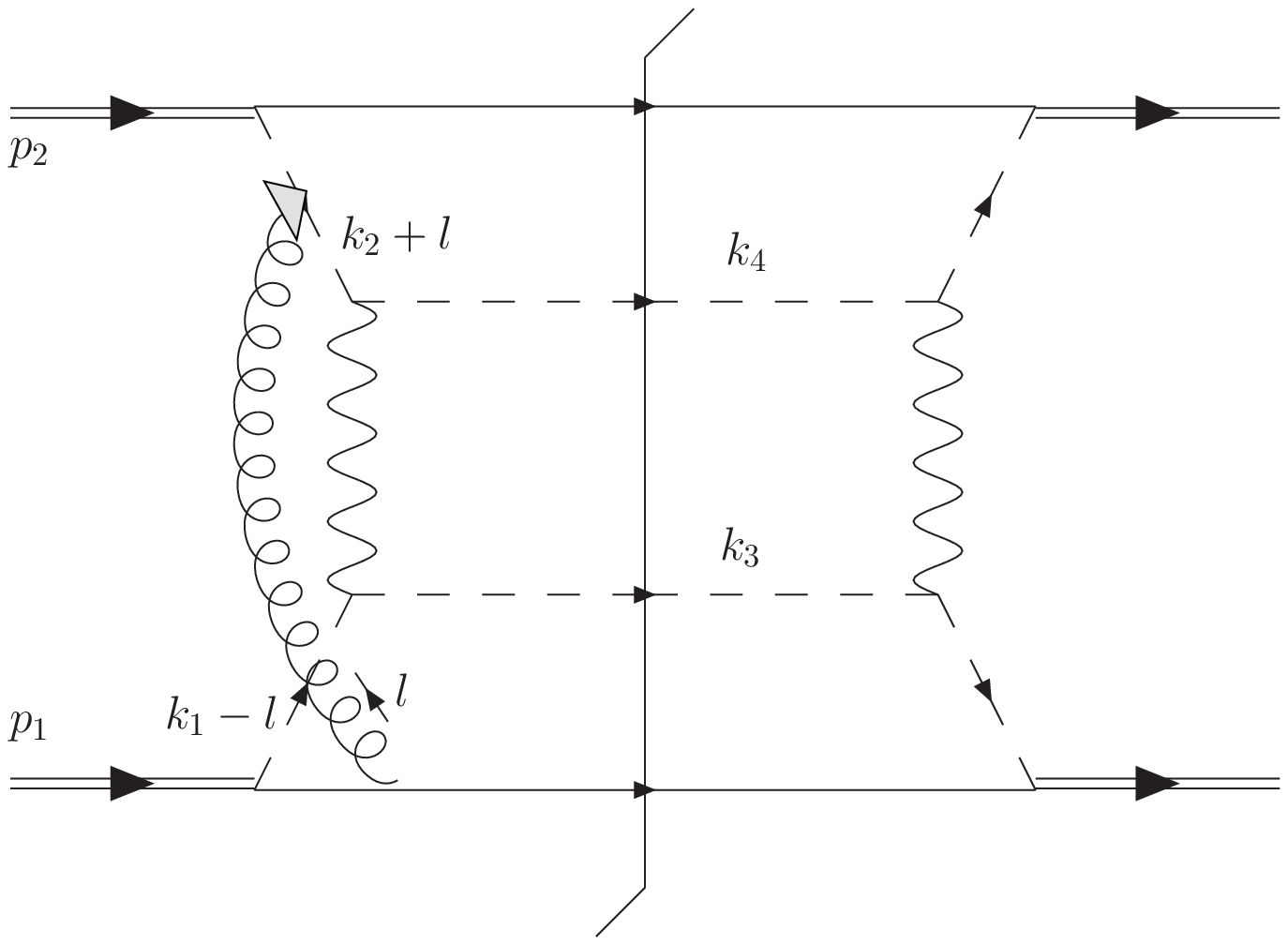}
  \\
  (a) & (b)
  \\[3mm]
    \includegraphics[scale=0.5]{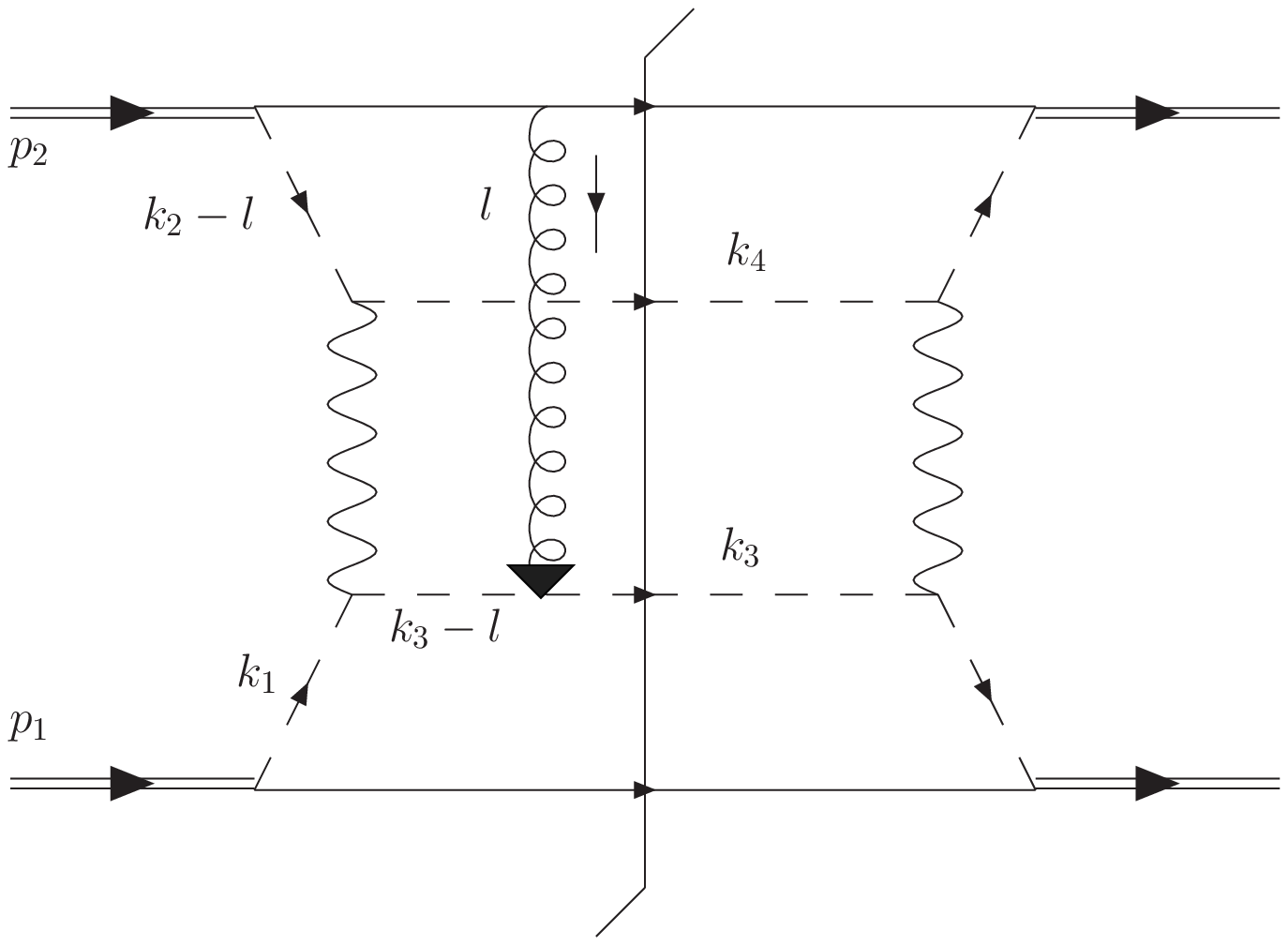}
    &
    \includegraphics[scale=0.5]{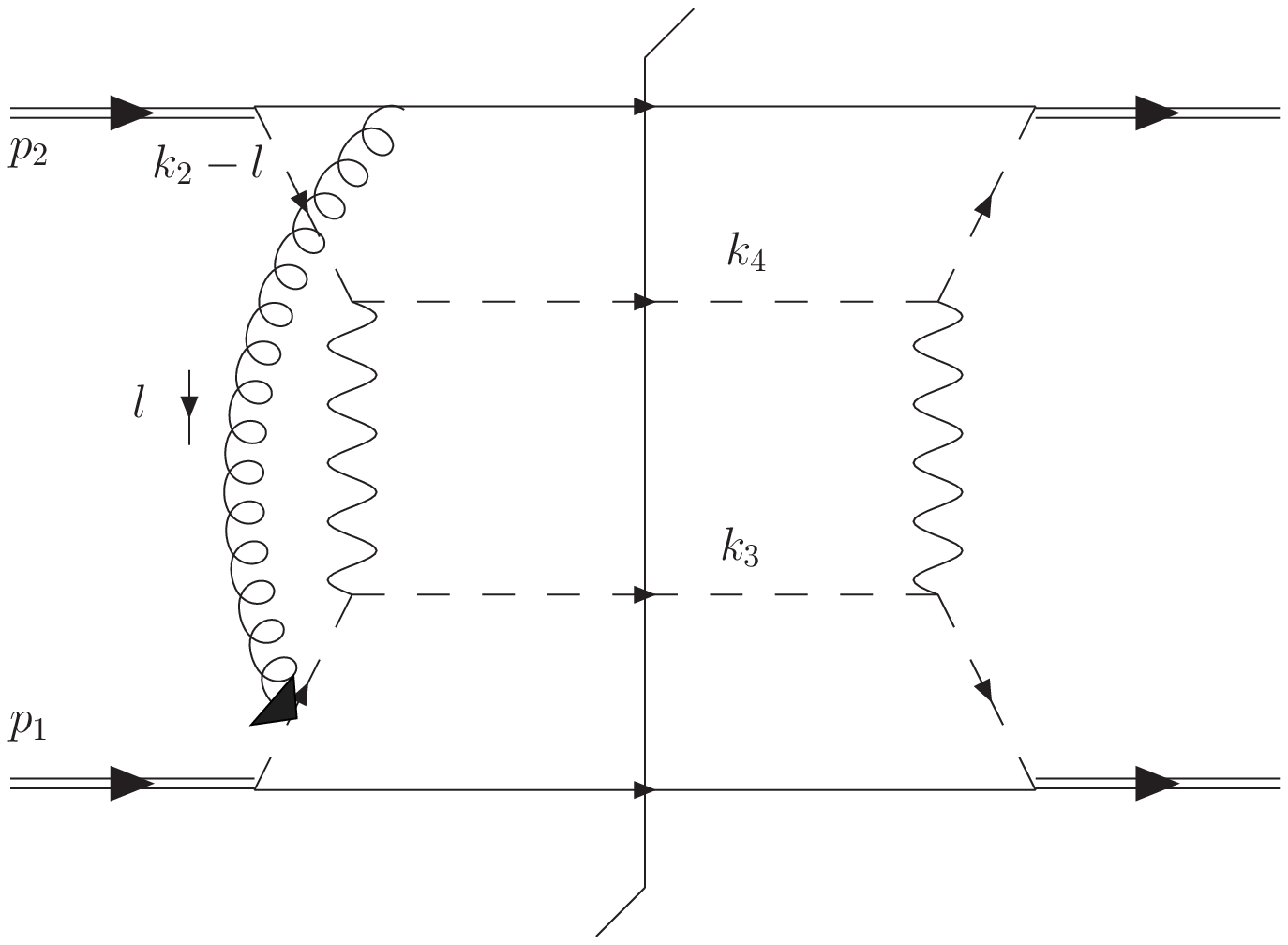}
   \\
   (c) & (d)
  \end{tabular}
\caption{Graphs of the type that led to a violation of TMD-factorization in Ref.~\cite{collins.qiu}. 
Gluons radiated from $H_1$ are illustrated in (a) and (b); gluons radiated from $H_2$ are illustrated in (c) and (d).
The Hermitian conjugate graphs should also be included.
These graphs vanish in our model because of their trivial color factor.}
\label{fig:oneglueviol}
\end{figure*}

Hence, the relevant types of graphs are represented by Figs.~\ref{fig:onegluestand} through~\ref{fig:violating}.
If a generalized TMD-factorization formula is possible, then the sum over all such graphs must produce a 
factorized form like Eq.~(\ref{eq:genfactor}) 
with a Wilson line structure in the TMD PDF or FF for each hadron separately.  
We will consider each type of graph in the following sections.

\section{One Extra Gluon}
\label{sec:onegluon}

We begin the investigation of diagrams by reviewing the steps for determining the contribution 
from a single extra gluon.  As in Ref.~\cite{collins.qiu}, we focus on the calculation of an SSA.
We start with graphs of the type shown in Fig.~\ref{fig:onegluestand}, where the extra gluon attaches on the side of the hard 
part nearest to its parent hadron.  
Any spin asymmetry disappears in the zeroth order cross section, Fig.~\ref{fig:basic}
because there are too few Dirac matrices to produce a non-zero result in the traces with $\gamma_5$.

Consider, for example, Fig.~\ref{fig:onegluestand}(a).  The arrow on the gluon line indicates that it is collinear to $H_1$.
By first deforming the $l$ integral out of the Glauber region to the $H_1$-collinear region, one may replace the intermediate struck quark line of momentum
$k_3 - l$ by the eikonal factor
\begin{equation}
\label{eq:oneeikonal}
\frac{t^a g n_1^\mu}{-l^+ + i \epsilon} =  - g t^a n_1^\mu {\rm P.V.} \frac{1}{l^+} - i g t^an_1^\mu \pi \, \delta(l^+)
\end{equation}
where $n_1^\mu \equiv (0,1,{\bf 0}_t)$.  The sign on the $i \epsilon$ is determined by the direction of the contour deformation.
For the spin-dependent part, the attachment of the extra gluon at the spectator produces a factor at leading power equal to
\begin{multline}
\label{eq:spin}
\frac{t^a}{2} \tr{ (\slashed{p}_1 + m_{H_1})  \gamma_5 \slashed{s}_1 (\slashed{p}_1 - \slashed{k}_1 + \slashed{l} + m_{\psi_1} ) 
\right. \times \\ \times \left. \gamma^+ (\slashed{p}_1 - \slashed{k}_1 + m_{\psi_1} ) } \\  
\approx 2 i t^a \epsilon_{jk} s_1^j l^k p^+ (m_{H_1} (1 - x_1) + m_{\psi_1}). 
\end{multline}
When this expression is combined with the imaginary part of Eq.~(\ref{eq:oneeikonal}), the factors of $-i$ and $i$ combine and a contribution 
to an SSA is obtained.  The $\epsilon_{jk}$ is the two-dimensional Levi-Civita symbol with $\epsilon_{12} = 1$. 

If the extra gluon is on the other side of the cut as in Fig.~\ref{fig:onegluestand}(b), the eikonal factor is
\begin{equation}
\label{eq:oneeikonal2}
\frac{t^a g n_1^\mu}{-l^+ - i \epsilon} =  - g t^a n_1^\mu {\rm P.V.} \frac{1}{l^+} + i g t^an_1^\mu \pi \, \delta(l^+).
\end{equation}
The factor from the attachment at the spectator is,
\begin{multline}
\label{eq:spin2}
\frac{t^a}{2} \tr{ (\slashed{p}_1 + m_{H_1})  
\gamma_5 \slashed{s}_1 (\slashed{p}_1 - \slashed{k}_1 + m_{\psi_1} ) \right. \times \\ \times \left. 
\gamma^+ (\slashed{p}_1 - \slashed{k}_1 + \slashed{l} + m_{\psi_1} ) } \\
\approx - 2 i t^a \epsilon_{jk} s_1^j l^k p^+ (m_{H_1} (1 - x_1) + m_{\psi_1}). 
\end{multline}
When this is combined with the imaginary part of Eq.~(\ref{eq:oneeikonal2}), the factors of $i$ and $-i$ 
combine and a contribution to an SSA is again obtained.
It is exactly equal to the one found in Fig.~\ref{fig:onegluestand}(a) as it must be since the graphs are related by 
Hermitian conjugation.  

Analogous steps apply to Figs.~\ref{fig:onegluestand}(c,d).  The different shadings on the arrows  
is to emphasize that they are radiated from $H_2$ rather than $H_1$.
Now, after a contour deformation $l$ can be made collinear to the minus direction, and a 
vector $n_2^\mu = (1,0,{\bf 0}_t)$ is used instead of $n_1^\mu$.
The eikonal factors are,
\begin{equation}
\label{eq:oneeikonalp}
\frac{t^a g n_2^\nu}{-l^- + i \epsilon} =  - g t^a n_2^\nu {\rm P.V.} \frac{1}{l^-} - i g t^a n_2^\nu \pi \, \delta(l^-)
\end{equation}
for Fig.~\ref{fig:onegluestand}(c) and
\begin{equation}
\label{eq:oneeikonal2p}
\frac{t^a g n_2^\nu}{-l^- - i \epsilon} =  - g t^a n_2^\nu {\rm P.V.} \frac{1}{l^-} + i g t^a n_2^\nu \pi \, \delta(l^-)
\end{equation}
for Fig.~\ref{fig:onegluestand}(d).

In our calculation of the SSA, $p_2$ is unpolarized so there is no 
factor of $i$ coming from the attachment to the upper spectator.  
Therefore, it is only the real principal 
value contributions that are kept in Eqs.~(\ref{eq:oneeikonalp},\ref{eq:oneeikonal2p}).
Note that there can be no \emph{double} Sivers effect at the level of one extra gluon because 
there must be at least one extra gluon from 
each of the hadrons for their PDFs to have spin dependence.

The $t^a$s in Eqs.~(\ref{eq:oneeikonal},\ref{eq:oneeikonal2})
combine with the $t^a$s in Eq.~(\ref{eq:spin},\ref{eq:spin2}) to produce an overall
color factor equal to
\begin{equation}
\label{eq:onecolor}
T_F (N_c^2 - 1).
\end{equation}
Exactly the same color factor also arises when $l$ is collinear to $p_2$ as in Figs.~\ref{fig:onegluestand}(c,d).

The remaining type of single-gluon graphs are those in Fig.~\ref{fig:oneglueviol}.
For the lines collinear to $H_1$ the eikonal propagators are 
\begin{multline}
\label{eq:oneviol}
\frac{\trc{t^a} g n_1^\mu}{-l^+ + i \epsilon} =  \\ - g \trc{t^a} n_1^\mu {\rm P.V.} \frac{1}{l^+} - i g \trc{t^a} n_1^\mu \pi \, \delta(l^+),
\end{multline}
from Fig.~\ref{fig:oneglueviol}(a) and 
\begin{multline}
\label{eq:oneviol2}
\frac{\trc{t^a} g n_1^\mu}{l^+ + i \epsilon} =  \\ + g \trc{t^a} n_1^\mu {\rm P.V.} \frac{1}{l^+} - i g \trc{t^a} n_1^\mu \pi \, \delta(l^+),
\end{multline}
from Fig.~\ref{fig:oneglueviol}(b).  
The sum of these graphs gives a contribution equal to
\begin{equation}
\label{eq:anomfact1}
- 2 i g \trc{t^a} \pi n_1^\mu \delta(l^+).
\end{equation}
If the color factor in Eq.~(\ref{eq:anomfact1}) did not vanish, then it would contribute to an SSA
when the factor of $i$ combines with the factor of $i$ in Eq.~(\ref{eq:spin}).
The same result is obtained from the Hermitian conjugate graphs.
In a totally unpolarized cross section, there would be no contribution from Eq.~(\ref{eq:anomfact1}), regardless of the color factor, since
it is imaginary.

A symmetric analysis applies if the extra gluon is instead radiated from hadron $H_2$ 
as in Figs.~\ref{fig:oneglueviol}(c,d).
In that case $n_1$ is replaced by $n_2$ and $l^+$ is replaced by $l^-$.  Since $p_2$ is unpolarized,
it is the real parts of the eikonal factors that contribute.  Thus, since the real parts cancel between graphs, the eikonal factors 
from Fig.~\ref{fig:oneglueviol}(c,d) do not contribute to the unpolarized TMD PDF of hadron $H_2$.

The uncanceled terms like Eq.~(\ref{eq:anomfact1}) would ordinarily 
signal a breakdown of standard TMD-factorization because they are not consistent with 
the standard Wilson line structure in a TMD PDF.
This is exactly what is observed for the Abelian gauge theory calculation in Ref.~\cite{collins.qiu}.
In our non-Abelian example, however, these single extra gluon contributions to a ``factorization anomaly'' are exactly 
zero because they happen to include the trivial color factor:
\begin{equation}
\label{eq:tracevanish}
\trc{t^a} = 0.
\end{equation}
So, in our specific non-Abelian calculation, Eqs.~(\ref{eq:oneeikonal},~\ref{eq:oneeikonal2}) are the only 
eikonal propagators that contribute to an SSA at the level of just one extra gluon.

The non-vanishing eikonal factors (coming from Figs.~\ref{fig:onegluestand}(a,b))
are exactly what is obtained from an order $g$ expansion of the 
Wilson lines in the \emph{standard} definition of the TMD PDFs:
\begin{multline}
\label{eq:standTMD}
\Phi_{H_1}^{[n_1]}(x_1,k_{1T}) = 
x_1 p_1^+ \int \frac{dw^- d^2 {\bf w}_t }{(2 \pi)^3} e^{ -i x_1 p_1^+ w^- + i {\bf k}_t \cdot {\bf w}_t} \times \\ \times
\langle H_1, s_1 | \phi_{1,j}^{\dagger}(0,w^-,{\bf w}_t) \, U^{[n_1]}_{jk}[0,w] \, \phi_{1,k}(0) | H_1, s_1 \rangle.
\end{multline}
The standard Wilson line operator 
$U^{[n_1]}_{jk}[0,w]$ is inserted between the two scalar quark fields.  The triplet color indices $j,k$ are shown explicitly to 
emphasize the flow of color.  
The full Wilson line insertion is,
\begin{equation}
\label{eq:fullwill}
U^{[n_1]}_{jk}[0,w] = \left[ V_{w}^{\dagger}(n_1) \right]_{jj^\prime} [I(n_1)]_{j^\prime k^\prime} \left[ V_{0}(n_1) \right]_{k^\prime k},
\end{equation}
where
\begin{multline}
\label{eq:wildef}
\left[ V_w (n_1) \right]_{j j^\prime} = \\ P \exp \left( 
-i g t^a \int_0^{\infty} d \lambda \, n_1 \cdot A^a(w + \lambda n_1 ) \right)_{j j^\prime},
\end{multline}
with $P$ denoting a path-ordering operator.
The superscript $[n_1]$ in Eq.~(\ref{eq:standTMD}) refers to the direction of the main leg of the Wilson line, starting from point $0$.
The extra gluon attachment in 
Fig.~\ref{fig:onegluestand}(a) on the
left side of the cut contributes to $\left[ V_{0}(n_1) \right]_{k^\prime k}$ while
the extra gluon in Fig.~\ref{fig:onegluestand}(b) contributes to $\left[ V_{w}^{\dagger}(n_1) \right]_{jj^\prime}$.
To close the Wilson line and ensure that it is exactly gauge invariant, the path needs a transverse link at light-cone infinity~\cite{Belitsky:2002sm},
\begin{equation}
\left[ I(n_1) \right]_{j j^\prime} = P \exp \left( -ig t^a  \int_C dz^{\mu} A^a_{\mu}(z) \right)_{j j^\prime},
\end{equation}
where $C$ is a path in the transverse direction connecting the points, $(0,\infty,{\bf w}_t)$ and $(0,\infty,{\bf 0}_t)$.
In a derivation of TMD-factorization in Feynman gauge, contributions to the link at infinity do not arise explicitly.
(See, e.g., Refs.~\cite{Collins:1989gx,Boer:2003cm,Collins:2008ht} and references therein for a review of the steps 
for resumming collinear gluon attachments and identifying the resulting gauge links.)

Note that the Wilson line insertion in the standard TMD PDF of Eq.~(\ref{eq:standTMD}) contracts the color indices of the quark fields, so 
$\Phi_{H_1}^{[n_1]}(x_1,k_{1T})$ has no leftover color indices --- it is a real-valued function as is appropriate for 
a gauge invariant probability density.

Similar steps result in the single-gluon contribution to 
the Wilson line in the standard TMD PDF for $H_2$ when the extra gluon 
is radiated from the upper spectator as in Figs.~\ref{fig:onegluestand}(c,d).
There the Wilson line insertion instead points in 
the direction $n_2 = (1,0, {\bf 0}_t)$:
\begin{multline}
\label{eq:standTMD2}
\Phi_{H_2}^{[n_2]}(x_2,k_{2T}) = 
x_2 p_2^- \int \frac{dw^+ d^2 {\bf w}_t }{(2 \pi)^3} e^{ -i x_2 p_2^- w^+ + i {\bf k}_t \cdot {\bf w}_t} \times \\ \times
\langle H_2, s_2 | \phi_{2,j}^{\dagger}(0,w^+,{\bf w}_t) \, U^{[n_2]}_{jk}[0,w] \, \phi_{2,k}(0) | H_2, s_2 \rangle.
\end{multline}

In the Abelian case considered in Ref.~\cite{collins.qiu}, breaking 
of standard TMD-factorization was observed for an SSA because the factorization 
anomaly terms analogous to Eq.~(\ref{eq:anomfact1}) did not vanish.  
There, the ``color factor'' was the Abelian charge $g_2$ for Abelian ``gluons'' 
rather than the non-Abelian color factor $\trc{t^a} = 0$.   
That the anomalous terms vanish at the level of 
one extra gluon in our non-Abelian example
is only due to the highly simple color structure 
in the particular hard process that we have considered.  
Generally, when color is exchanged in the 
hard part, standard TMD-factorization breaking
will already appear at the level of one extra gluon.  
In our example, violations of standard TMD-factorization only appear at the level 
of two extra gluons or higher. 

Schematically, the standard TMD-factorization formula suggested by the sum of one-extra-gluon graphs in Figs.~\ref{fig:onegluestand}(a-d) is
\begin{multline}
\label{eq:standfact}
d \sigma \stackrel{!}{=} \mathcal{H} \otimes \Phi_{H_1}^{[n_1]}(x_1,k_{1T})  \otimes \Phi_{H_2}^{[n_2]}(x_2,k_{2T}) 
\otimes \\ \otimes \delta^{(2)}({\bf k}_{1T} + {\bf k}_{2T} - {\bf k}_{3T} - {\bf k}_{4T}).
\end{multline}
Here $\mathcal{H}$ is the same hard factor that appeared at zeroth order in Sect.~\ref{sec:setup}.
The ``$!$'' on the equal sign is to emphasize that this formula is 
ultimately incorrect, as explained in Refs.~\cite{collins.qiu,Collins:2007jp}.  
We will see this explicitly at the level two extra gluons in the next section.

\section{Two Extra Gluons}
\label{sec:twogluons}

\begin{figure*}
\centering
\includegraphics[scale=.5]{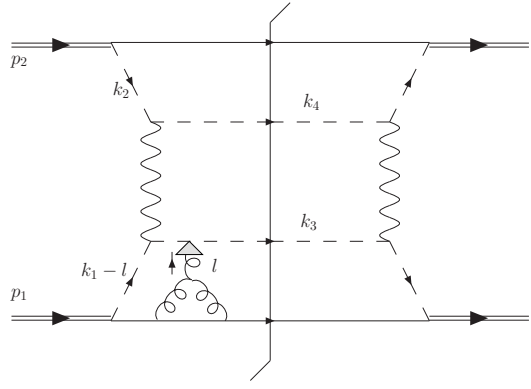}
\caption{A gluon three-point interaction that contributes to the single gluon Wilson line attachment.}
\label{fig:triple}
\end{figure*}
\begin{figure*}
\centering
  \begin{tabular}{c@{\hspace*{5mm}}c}
    \includegraphics[scale=0.5]{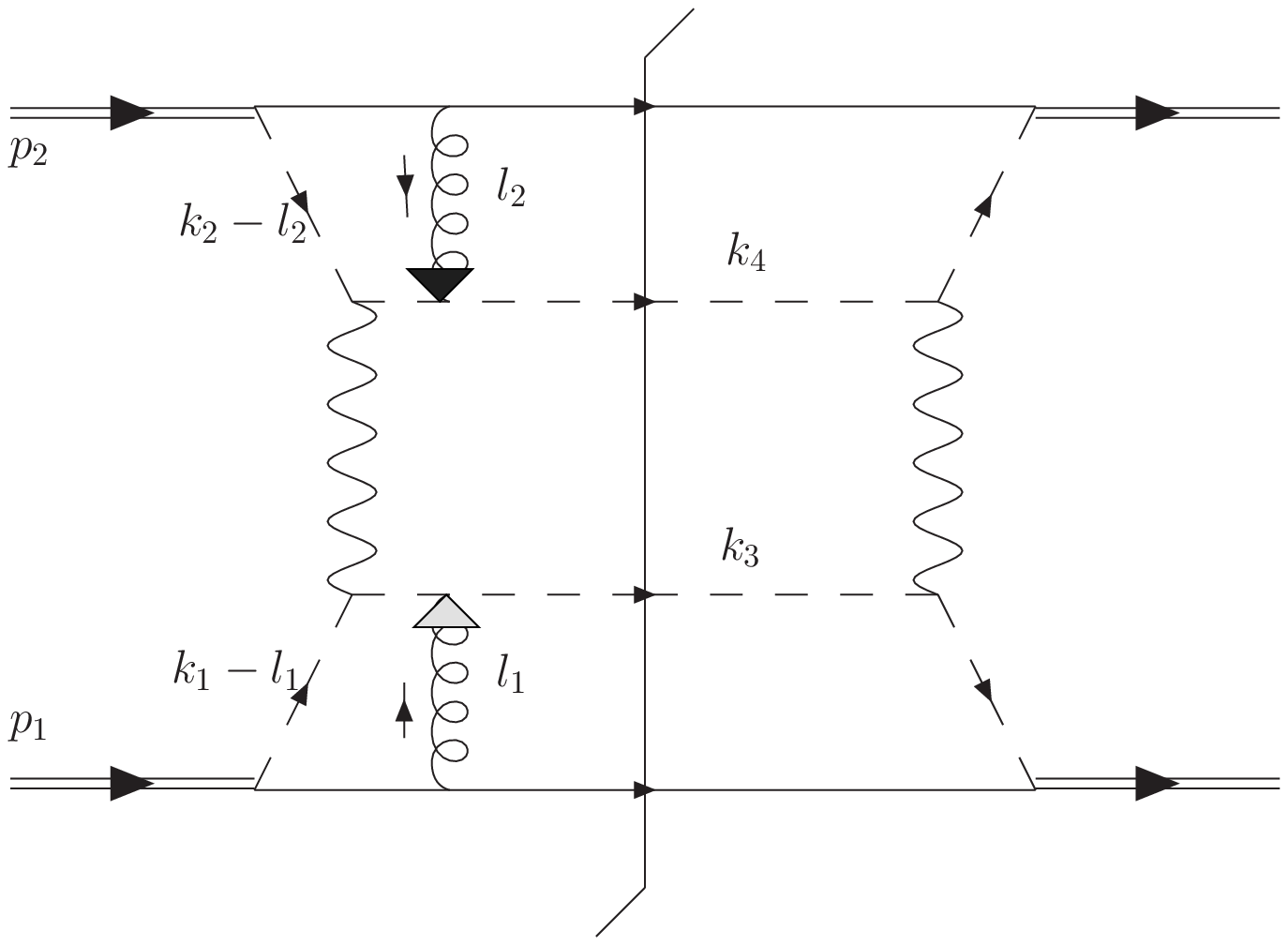}
    &
    \includegraphics[scale=0.5]{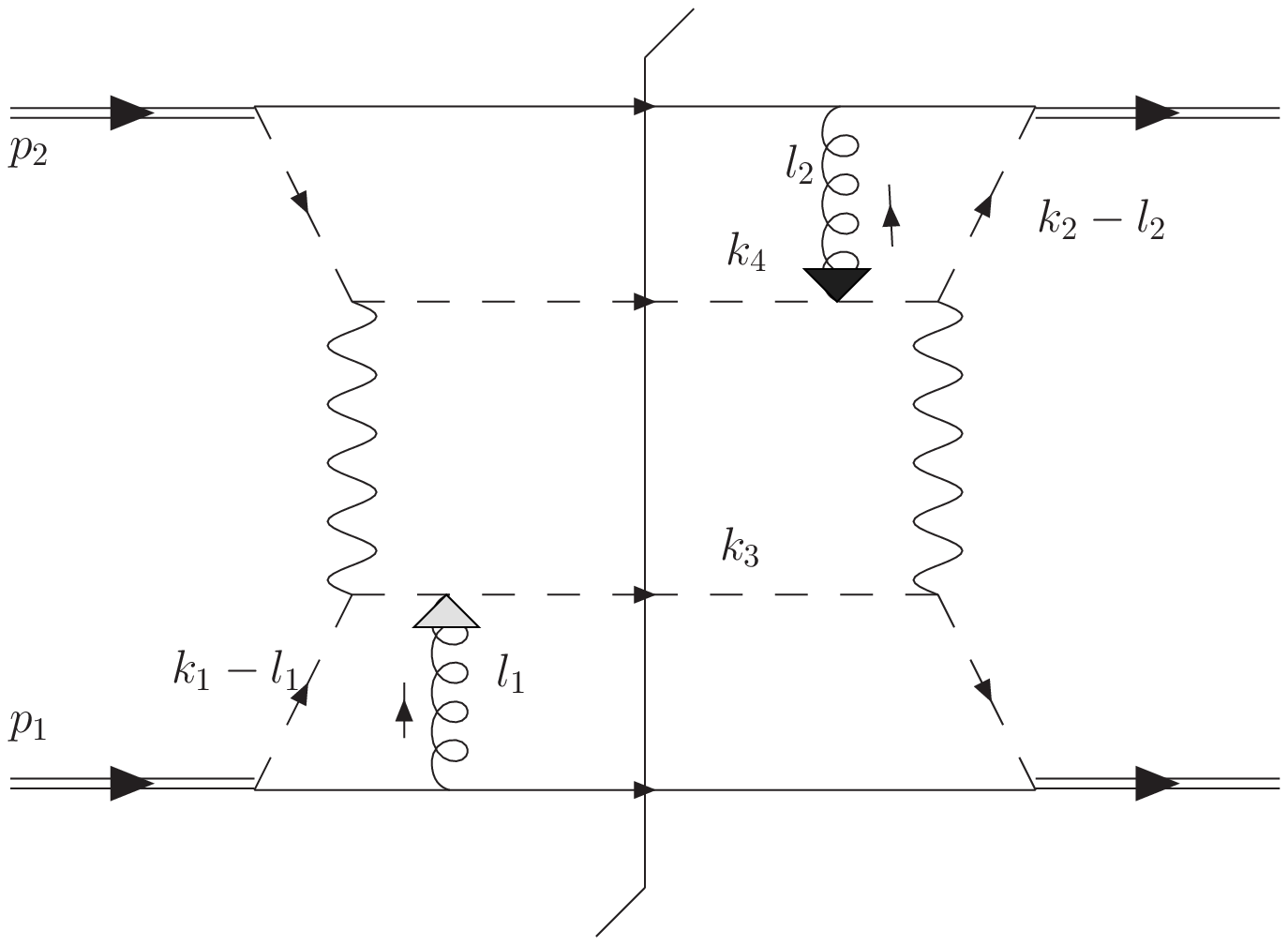}
  \\
  (a) & (b)
  \\[3mm]
    \includegraphics[scale=0.5]{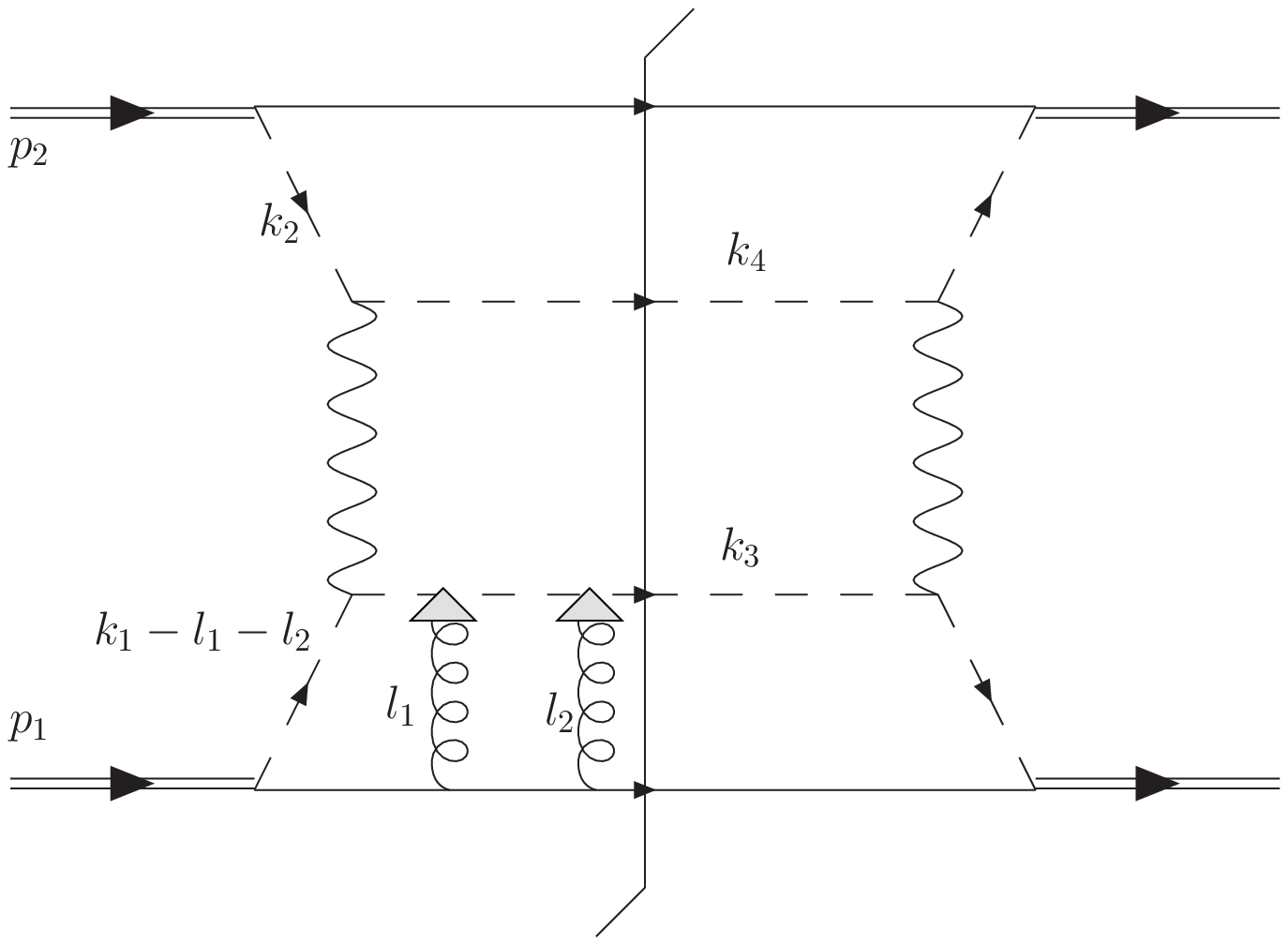}
    &
    \includegraphics[scale=0.5]{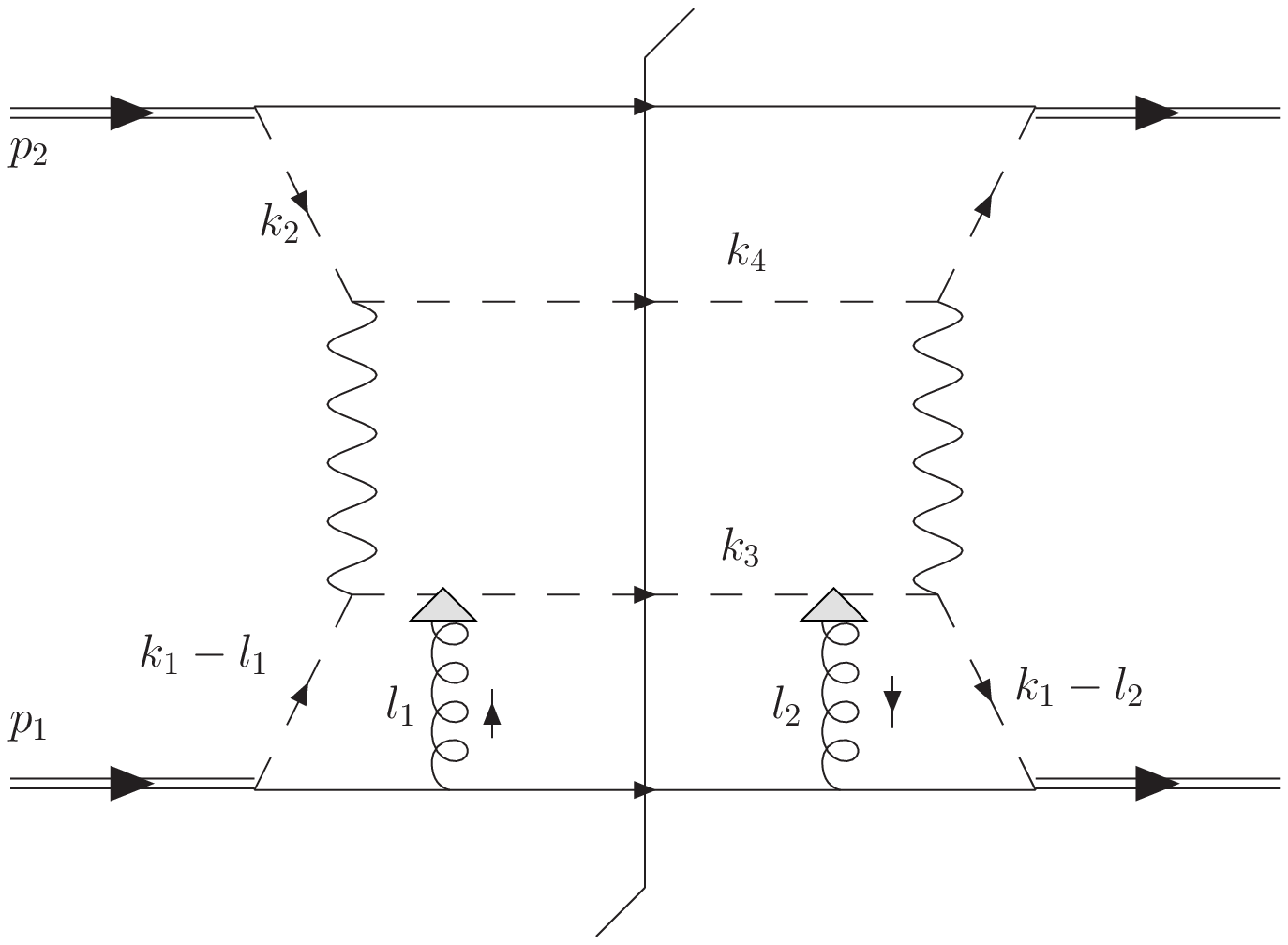}
   \\
   (c) & (d)
  \end{tabular}
\caption{Graphs that contribute to the standard Wilson lines at the level of two extra gluons.  The Hermitian conjugate graphs should also be included, as 
well as graphs with both gluons collinear to $p_2$.}
\label{fig:twoopp}
\end{figure*}
\begin{figure*}
\centering
  \begin{tabular}{c@{\hspace*{5mm}}c}
    \includegraphics[scale=0.5]{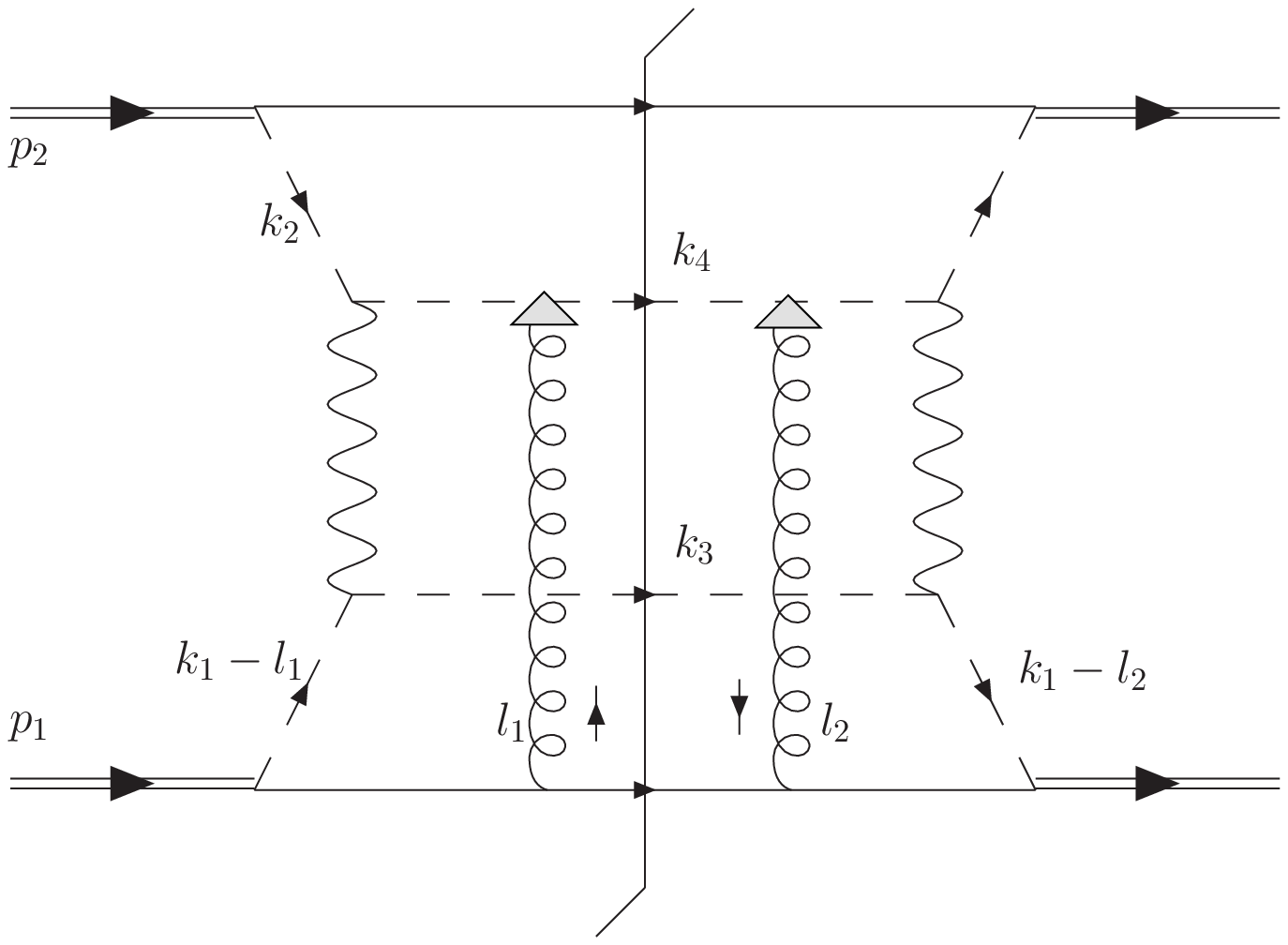}
    &
    \includegraphics[scale=0.5]{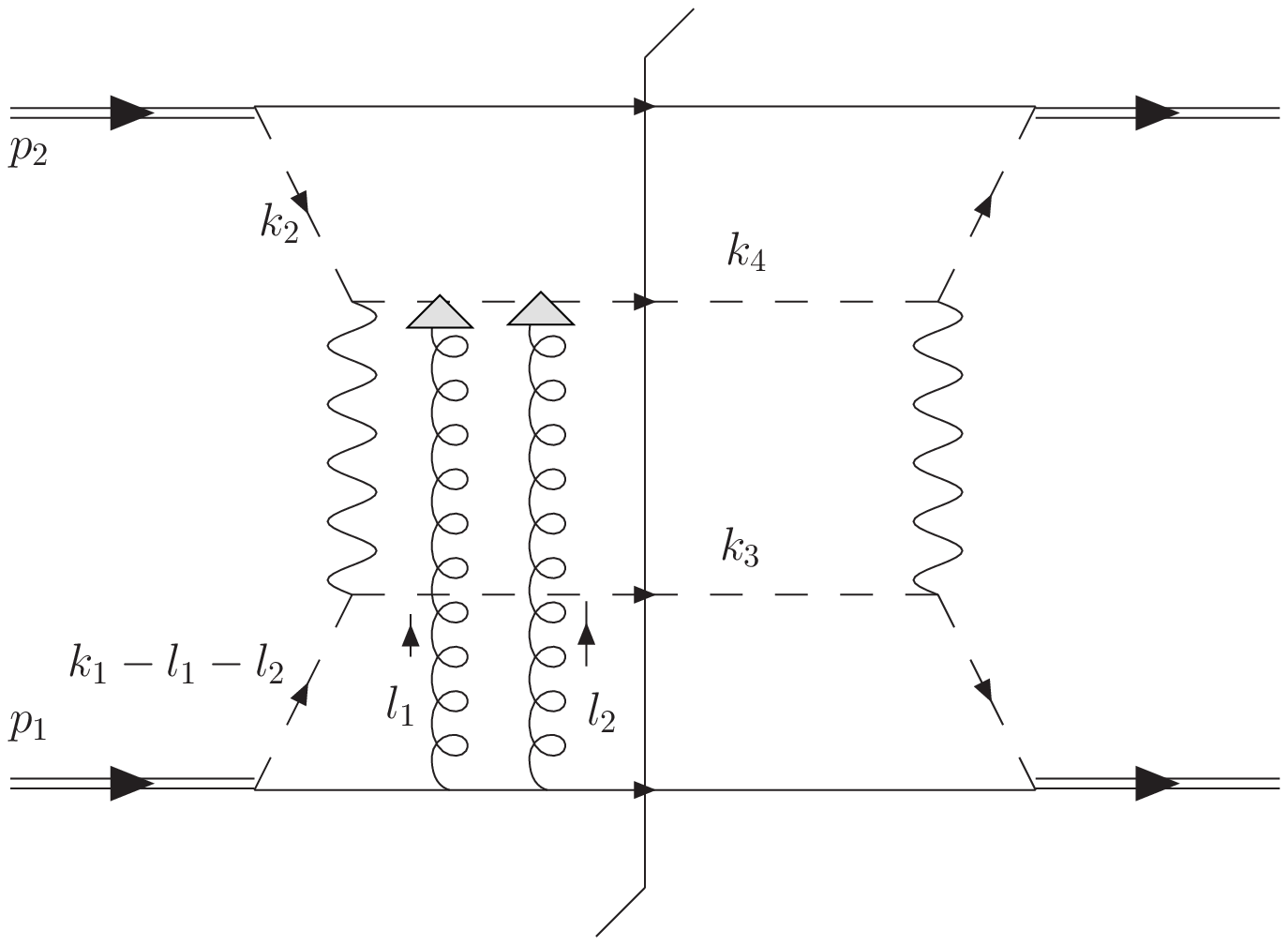}
  \\
  (a) & (b)
  \end{tabular}
\caption{Graphs that contribute to a factorization anomaly.  Other graphs 
that should be included are all other graphs with both gluons attaching the lower spectator to 
the $k_2$ and $k_4$ lines at the upper part of the graph.
In addition, the anomalous contribution to the TMD PDF of $H_2$ is 
obtained by including the graphs with both gluons radiated from $p_2$ and attaching the upper spectator 
to the $k_1$ and $k_3$ lines.  There is a total of $32$ graphs of this type, including graphs with both gluons radiated from $p_2$.}
\label{fig:twosame}
\end{figure*}
In this section we review the steps for illustrating 
a violation of standard TMD-factorization at the level of two 
extra gluons radiated from \emph{one} of the hadrons.  The 
structure of the resulting TMD-factorization anomaly terms  
constrains the possible generalizations of TMD-factorization, as we will discuss more in Sect.~\ref{sec:genfact}.

The simple color structure of the hard process strongly limits the 
number of non-vanishing graphs.  
All graphs in which the second gluon is internal to the hadron subgraph (e.g. Fig.~\ref{fig:triple}) 
only give further contributions to a single gluon contribution in the \emph{standard} Wilson lines in Eq.~(\ref{eq:standfact}).
The only non-vanishing graphs that give order $g^2$ contributions to a Wilson line operator are of the 
type shown in Figs.~\ref{fig:twoopp},~\ref{fig:twosame},~\ref{fig:violating}. 
All other graphs vanish because they 
include an overall color factor $\trc{t^a} = 0$.

We will consider each type of graph in turn in the next few subsections.
We will first identify the graphs that are consistent with the 
standard TMD-factorization formula Eq.~(\ref{eq:standfact}). 
Next we will consider the graphs that contribute to a violation 
of standard TMD-factorization that were already discussed in Refs.~\cite{Collins:2007jp}.
The remaining graphs in Fig.~\ref{fig:violating} are ultimately responsible for breaking 
generalized factorization, so we will put off any further discussion of them until Sect.~\ref{sec:breakdown}.

\subsection{Graphs Contributing to Standard Factorization}
 
By repeating the steps of Sect.~\ref{sec:onegluon},
graphs of the type shown in Figs.~\ref{fig:twoopp}  
lead to the Wilson lines in Eqs.~(\ref{eq:standTMD},~\ref{eq:standTMD2}) for the TMD PDFs of  
the standard TMD-factorization formula.  
The eikonal factors that arise from Figs.~\ref{fig:twoopp}(a,b) 
correspond to keeping the order $g$ Wilson line
contribution from each of the TMD PDFs individually. 
Figures~\ref{fig:twoopp}(c,d) (and related graphs) 
correspond to keeping the order $g^2$ Wilson line 
contributions in one of the TMD PDFs, and the zeroth order from the other.

The graphs considered up to now exhaust the 
contributions obtained by expanding each of the Wilson lines in 
the \emph{standard} TMD-factorization formula Eq.~(\ref{eq:standfact}) up to order $g^2$ (in the Wilson line).  
Therefore, any 
remaining uncanceled contributions from the 
graphs of the type shown in Figs.~\ref{fig:twosame},~\ref{fig:violating}
violate the standard TMD-factorization formula.

\subsection{Violation of Standard TMD-Factorization}
\label{sec:stanvio}

The graphs of the type shown in Fig.~\ref{fig:twosame}, where two gluons 
attach the spectator of \emph{one hadron} to the 
opposite side of the hard subgraph, have been shown in 
Refs.~\cite{Collins:2007jp,Vogelsang:2007jk} to be non-vanishing in  
calculations of both the unpolarized cross section and an SSA for an Abelian gauge theory.  
These graphs, therefore, yield a violation of the 
standard TMD-factorization formula Eq.~(\ref{eq:standfact}). 

The contribution to an anomalous SSA requires an imaginary part from the 
extra eikonal factors. 
In Ref.~\cite{Collins:2007jp} the imaginary part came from graphs with one gluon 
attaching near to the parent hadron and the other attaching  
near to the opposite hadron (e.g., Fig.~\ref{fig:zero}(b)).  These graphs vanish in our 
model because they again include a color factor $\trc{t^a} = 0$.

The non-cancellation of the unpolarized factorization anomaly, 
however, remains.  It arises from graphs with both extra gluons radiating from the same hadron and  
attaching at the side of the hard subgraph near to the opposite hadron, as in Fig.~\ref{fig:twosame}.
To make the violation of standard TMD-factorization explicit, we will briefly review the steps of Ref.~\cite{Collins:2007jp}.
We are for the moment only concerned with the unpolarized cross section so we will temporarily 
simplify the model of Sect.~\ref{sec:setup} by making all the hadron and spectator fields scalars.
We further simplify the calculation by assuming 
$m_{q_1} = m_{q_2} = m_{\psi_1} = m_{\psi_2} = m_q$.

For the graphs with the extra gluons on opposite 
sides of the cut (Fig.~\ref{fig:twosame}(a) and related graphs), the factorization 
anomaly for the unpolarized cross section comes from 
replacing the extra off-shell quark propagators by 
their eikonal propagators.  
Including all ways of attaching $l_1$ and $l_2$ to the $k_2$ and $k_4$ lines,
one finds the same result as in Ref.~\cite{Collins:2007jp}, but 
with the Abelian $g_1$ and $g_2$ charges replaced by the appropriate color factor:
\begin{multline}
\label{eq:anom1}
\trc{t^a t^b} g^2  n_1^\mu n_1^\nu \left( \frac{1}{-l_1^+ + i \epsilon} + \frac{1}{l_1^+ + i \epsilon} \right)  \times \\ \times 
\left(  \frac{1}{-l_2^+ - i \epsilon} + \frac{1}{l_2^+ - i \epsilon} \right)  = \\ 
4 \pi^2 g^2 n_1^\mu n_1^\nu \trc{t^a t^b} \delta(l_1^+) \delta(l_2^+).
\end{multline}

For the graphs with both extra gluons coupling on the same side of the cut 
(Fig.~\ref{fig:twosame}(b) and related graphs)), the steps are similar.   
Again including all ways of attaching $l_1$ and $l_2$ at the upper part of the graph, one finds
\begin{widetext}
\begin{multline}
\label{eq:anom2}
\trc{t^a t^b} g^2  n_1^\mu n_1^\nu \left\{ \left( \frac{1}{-l_1^+ + i \epsilon} \right) \left( \frac{1}{- l_2^+ + i \epsilon} \right) 
+ \left( \frac{1}{l_1^+ + i \epsilon} \right) \left( \frac{1}{- l_2^+ + i \epsilon} \right) + \right. \\ \left. +
\left( \frac{1}{-l_1^+ + i \epsilon} \right) \left( \frac{1}{l_2^+ + i \epsilon} \right) 
+ \left( \frac{1}{l_1^+ + i \epsilon} \right) \left( \frac{1}{l_2^+ + i \epsilon} \right) \right\} \\ 
= \trc{t^a t^b} g^2  n_1^\mu n_1^\nu \left( \frac{1}{l_1^+ + i \epsilon} + \frac{1}{-l_1^+ + i \epsilon} \right) 
\left( \frac{1}{l_2^+ + i \epsilon} + \frac{1}{-l_2^+ + i \epsilon} \right) 
= - 4 \pi^2 g^2 n_1^\mu n_1^\nu \trc{t^a t^b} \delta(l_1^+) \delta(l_2^+).
\end{multline}
The anomalous eikonal factor here is the
same as in Eq.~(\ref{eq:anom1}), apart from an overall minus-sign.
However, the propagator denominators for graphs like Fig.~\ref{fig:twosame}(a) (both gluons on opposite sides of the cut) 
are different from the propagator denominators for graphs like Fig.~\ref{fig:twosame}(b)
(gluons on the same side).  
For graphs with the extra gluons on opposite sides of the cut, one finds the following contribution to the 
TMD PDF of hadron $H_1$:
\begin{multline}
\label{eq:I1}
I_1(k_{1T}) = \\ \frac{g^2 \lambda_1^2 \trc{t^a t^b} \trc{t^b t^a} }{(2 \pi)^{12}} x_1 p_1^+ \int d k^- d^4 l_1 d^4 l_2 
\frac{ \left[ 2(p_1^+ - k_1^+) + l_1^+ \right] 
\left[ 2(p_1^+ - k_1^+) + l_2^+ \right] }{ (l_1^2 + i \epsilon)(l_2^2  - i \epsilon) 
\left[ (k_1 - l_1)^2 - m_q^2 + i \epsilon \right] \left[ (k_1 - l_2)^2 - m_q^2 - i \epsilon \right]} \times \\ 
\times
\frac{(2 \pi)^3 \delta(l_1^+) \delta(l_2^+) \delta((p_1 - k_1)^2 - m_q^2)}{\left[ (p_1 - k_1 + l_1)^2 - m_q^2 + i \epsilon \right] 
\left[ (p - k + l_2)^2 - m_q^2 - i \epsilon \right]}\\
= \frac{g^2 \lambda_1^2 T_F^2 (N_c^2 - 1) x_1 (1 - x_1) }{256 \pi^7} \int d^2 {\bf l}_{1T} \, d^2 {\bf l}_{2T} \prod_{j = 1,2} \frac{1}{l_{jT}^2 \left[ 
({\bf k}_{1T} - {\bf l}_{jT})^2 + m_q^2\right]}. 
\end{multline}
This is the same result as in Ref.~\cite{Collins:2007jp}, except that the gluon 
is massless and there is a non-Abelian color factor multiplying the integral.
Equation~(\ref{eq:anom2}) allows for a similar calculation of 
the remaining contribution to the TMD PDF for hadron $H_1$ from the graphs with the extra gluons on the same side of the cut:  
\begin{equation}
\label{eq:I2}
I_2(k_{1T}) 
= \frac{-g^2 \lambda_1^2 T_F^2 (N_c^2 - 1) x_1 (1 - x_1) }{256 \pi^7} \int d^2 {\bf l}_{1T} \, d^2 {\bf l}_{2T} \frac{1}{l_{1T}^2 l_{2T}^2 
\left[ ( {\bf k}_{1T} - {\bf l}_{1T} - {\bf l}_{2T})^2 + m_q^2 \right] \left[ k_{1T}^2 + m_q^2 \right]}. 
\end{equation}
\end{widetext}
The mismatch in denominators between Eq.~(\ref{eq:I1}) and Eq.~(\ref{eq:I2}) means that the 
full contribution $I_1(k_{1T}) + I_2(k_{1T})$ does not generally vanish point-by-point in $k_{1T}$.  
The sum of graphs like Fig.~\ref{fig:twosame} therefore results in uncanceled terms that 
are not accounted for by the standard Wilson lines.
Hence, the standard TMD-factorization formula Eq.~(\ref{eq:standfact}) fails for unpolarized scattering. 

Exactly analogous observations apply to the TMD PDF of the other hadron if the 
two extra gluons are radiated from the spectator in $H_2$, collinear to the minus direction.
In that case, the eikonal factors analogous to 
Eqs.~(\ref{eq:anom1},~\ref{eq:anom2}) will instead use a vector $n_2$ and 
the delta functions from the eikonal factors will be $\delta(l_1^-)$ and $\delta(l_2^-)$~\footnote{In earlier sections we 
neglected gluon exchanges between active quarks because they do not lead to spin dependence.  In the unpolarized cross 
section, the complete result is obtained only after including interactions between active quarks.  In the collinear
regions they result in the same set of eikonal factors as in Eqs.~(\ref{eq:anom1},~\ref{eq:anom2}) and analogous 
Wilson line contributions.  The key point is that the usual steps that normally would  
lead to a cancellation of the factorization anomaly between different cuts of the same graph in collinear factorization, fails in the transverse 
momentum dependent case.}.  So, both TMD PDFs yield factorization anomalies at the two-gluon level in the unpolarized cross section.

\section{Generalized TMD-factorization}
\label{sec:genfact}

From the factorization anomaly terms Eqs.~(\ref{eq:anom1},~\ref{eq:anom2}) one may determine 
what the modified gauge link structure must be for the TMD PDF of hadron $H_1$ in a generalized 
factorization formula~\cite{Bomhof:2004aw,Bomhof:2006dp,Bomhof:2007xt}.
In our example, the sequence of eikonal factors in the 
factorization anomaly terms of Eqs.~(\ref{eq:anom1},\ref{eq:anom2}), 
including the trace around a color loop, require a color-traced Wilson loop operator to be inserted 
into the definition, Eq.~(\ref{eq:standTMD}), 
of the TMD PDF for $H_1$.   
Each of the eikonal factors in Eqs.~(\ref{eq:anom1},\ref{eq:anom2}) corresponds to an attachment to a
leg of the Wilson loop.
Therefore, 
the standard TMD PDF in Eq.~(\ref{eq:standfact}) for $H_1$ should be replaced with, 
\begin{multline}
\label{eq:modTMD}
\Phi_{H_1}^{[n_1, (\Box) ]}(x_1,k_{1T}) = \\
x_1 p_1^+ \int \frac{dw^- d^2 {\bf w}_t }{(2 \pi)^3} e^{-i x_1 p_1^+ w^- + i {\bf k}_t \cdot {\bf w}_t }\times \\ \times
\langle H_1, s_1 | \phi_{1,r}^{\dagger}(0,w^-,{\bf w}_t) \, U^{n_1}_{rs}[0,w] U^{n_1}_{(\Box)} \, \phi_{1,s}(0) | H_1, s_1 \rangle.
\end{multline}
This is the same as the standard TMD PDF definition in Eq.~(\ref{eq:standTMD}) apart from the insertion 
of the following color-traced Wilson loop operator:
\begin{multline}
\label{eq:loop}
U^{n_1}_{(\Box)} = U^{n_1}_{ij}[0,w] \, (U^{n_1 \, \dagger}[0,w])_{ji} 
= \\ \trc{  V_0(n_1)   I(n_1) V_{w}^{\dagger}(n_1) 
V_{w}(n_1) I^{\dagger}(n_1) V_0^{\dagger}(n_1)} . 
\end{multline}
The sequence of eikonal factors in Eqs.~(\ref{eq:anom1},\ref{eq:anom2}) correspond exactly to gluons 
attaching to the ``$V$'' Wilson lines in Eq.~(\ref{eq:loop}). 
As before, the transverse links at infinity do not contribute to leading power in Feynman gauge.
The color trace corresponds to the trace over the color loop in the upper part of the 
graphs in Fig.~\ref{fig:twosame} and is what is needed to get the factor of $\trc{t^a t^b}$ in Eqs.~(\ref{eq:anom1},\ref{eq:anom2}).
Hence, the terms that violate standard factorization 
arise from expanding the $U^{n_1}_{(\Box)}$ operator in 
Eq.~(\ref{eq:modTMD}) up to order $g^2$ (see also Ref.~\cite{Vogelsang:2007jk}).

An exactly analogous TMD PDF is obtained for hadron $H_2$ when both extra gluons 
are radiated from $H_2$ in the collinear minus direction and attach to the lower half of the graph.
The main legs of the Wilson loop in the TMD PDF for $H_2$ point in the 
direction of a vector $n_2$:
\begin{multline}
\label{eq:modTMD2}
\Phi_{H_2}^{[n_2, (\Box) ]}(x_2,k_{2T}) = \\
x_2 p_2^- \int \frac{dw^+ d^2 {\bf w}_t }{(2 \pi)^3} e^{-i x_2 p_1^- w^+ + i {\bf k}_t \cdot {\bf w}_t }\times \\ \times
\langle H_2, s_2 | \phi_{2,r}^{\dagger}(0,w^+,{\bf w}_t) \, U^{n_2}_{rs}[0,w] U^{n_2}_{(\Box)} \, \phi_{2,s}(0) | H_2, s_2 \rangle.
\end{multline}

Thus, the factorization anomaly terms from Sect.~\ref{sec:stanvio} specify the type of new Wilson line insertions
that are needed if generalized TMD-factorization is to hold.
The \emph{generalized} TMD-factorization formula is:
\begin{multline}
\label{eq:modfact}
d \sigma \stackrel{!}{=} \mathcal{H} \otimes \Phi_{H_1}^{[n_1,(\Box)]}(x_1,k_{1T})  \otimes \Phi_{H_2}^{[n_2,(\Box)]}(x_2,k_{2T}) \otimes 
\\ \otimes \delta^{(2)}({\bf k}_{1T} + {\bf k}_{2T} - {\bf k}_{3T} - {\bf k}_{4T}).
\end{multline}
Again, $\mathcal{H}$ is the same hard factor that appeared at zeroth order in Sect.~\ref{sec:setup}.
The only difference from Eq.~(\ref{eq:standfact}) comes from the $U^{n}_{(\Box)}$ insertions in the TMD PDFs, indicated by the $(\Box)$ superscripts
on the individual TMD PDFs in Eq.~(\ref{eq:modfact}). 
We have again included a ``$!$'' over the equal sign to indicate that even this formula will ultimately fail when the remaining graphs are considered.

Equation~(\ref{eq:modfact}), with the TMD PDFs defined as in Eq.~(\ref{eq:modTMD}), 
follows the method proposed in~\cite{Bomhof:2004aw,Bomhof:2006dp,Bomhof:2007xt},
and is 
the most natural generalization of TMD-factorization.
For the purpose of our counter-argument to generalized TMD-factorization, the crucial 
point is that the new Wilson line insertion Eq.~(\ref{eq:loop}) involves a trace over color.
If we instead tried contracting the color indices of the new Wilson lines with the quark 
fields at points $0$ and $\xi$, then we would obtain the wrong color factor
for the anomalous 
contribution in Eqs.~(\ref{eq:anom1},\ref{eq:anom2}).
An even more serious problem would be that the generalized TMD-factorization formula would 
produce spurious terms at the \emph{one}-gluon level, already treated in Sect.~\ref{sec:onegluon}.
Namely, single-gluon attachments to the extra legs of the Wilson line would 
introduce factorization anomaly contributions analogous to Eq.~(\ref{eq:anomfact1}), 
but with different non-vanishing color factors, 
contradicting the analysis from Sect.~\ref{sec:onegluon} which found \emph{no} violation of standard TMD-factorization 
at order $g$ in the expansion of the Wilson lines (polarized or unpolarized).
Thus, consistency between the order-$g$ and order-$g^2$ contributions to the Wilson lines 
requires that the new Wilson line structures include the trace over color as in Eq.~(\ref{eq:loop}).

Other possible redefinitions of the TMD PDFs are unrelated to the color structure of the Wilson lines in the main correlation functions.
These include the modifications mentioned in Sect.~\ref{sec:setup}, such as tilting the direction of the Wilson line slightly away from the light-like
direction.
Thus, accounting for all contributions up to order $g^2$, in the Wilson line, and requiring consistency
between different orders in $g$, strongly constrains the possibilities for a generalization of TMD-factorization.
\begin{figure*}
\centering
  \begin{tabular}{c@{\hspace*{5mm}}c}
    \includegraphics[scale=0.5]{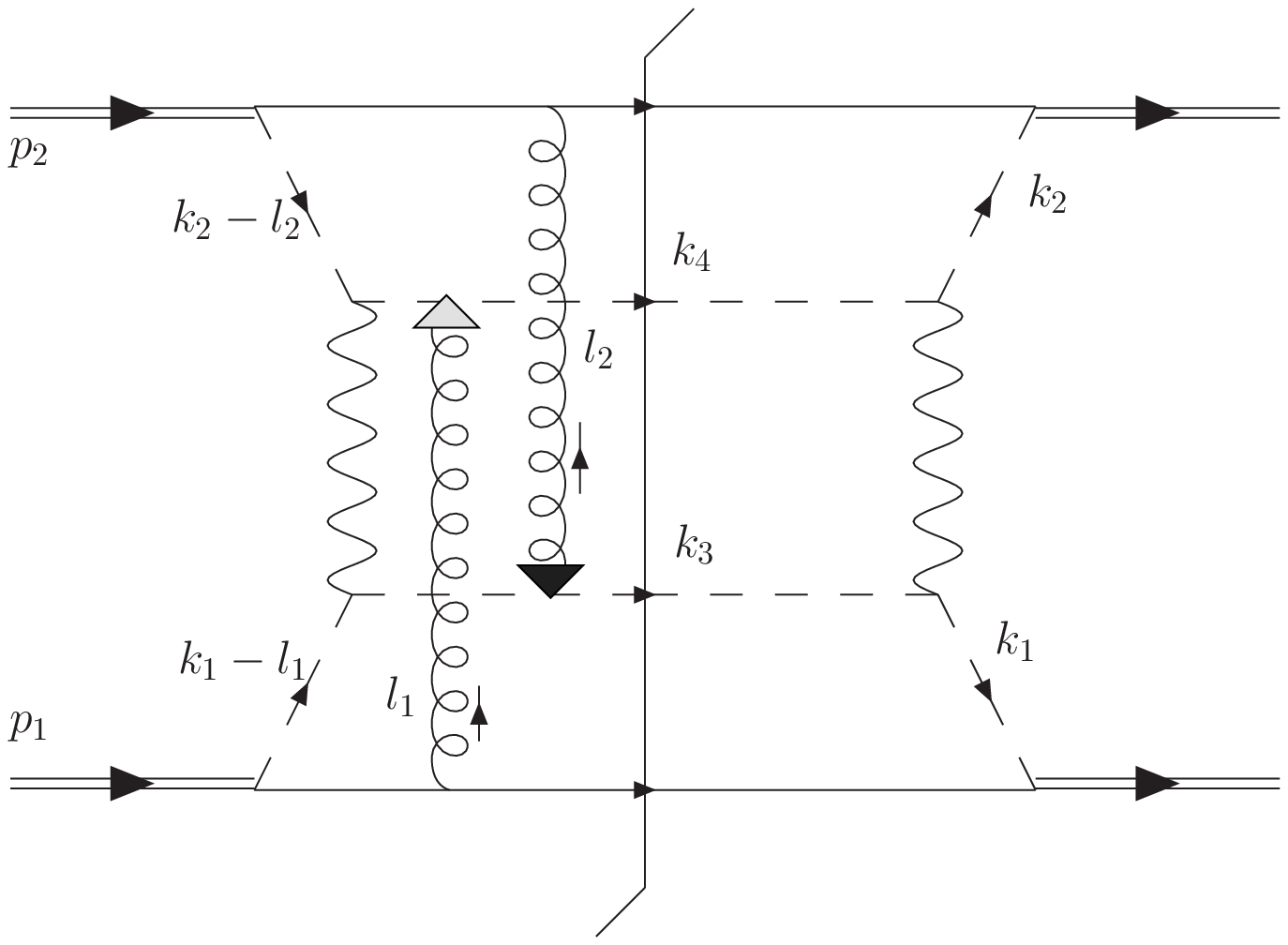}
    &
    \includegraphics[scale=0.5]{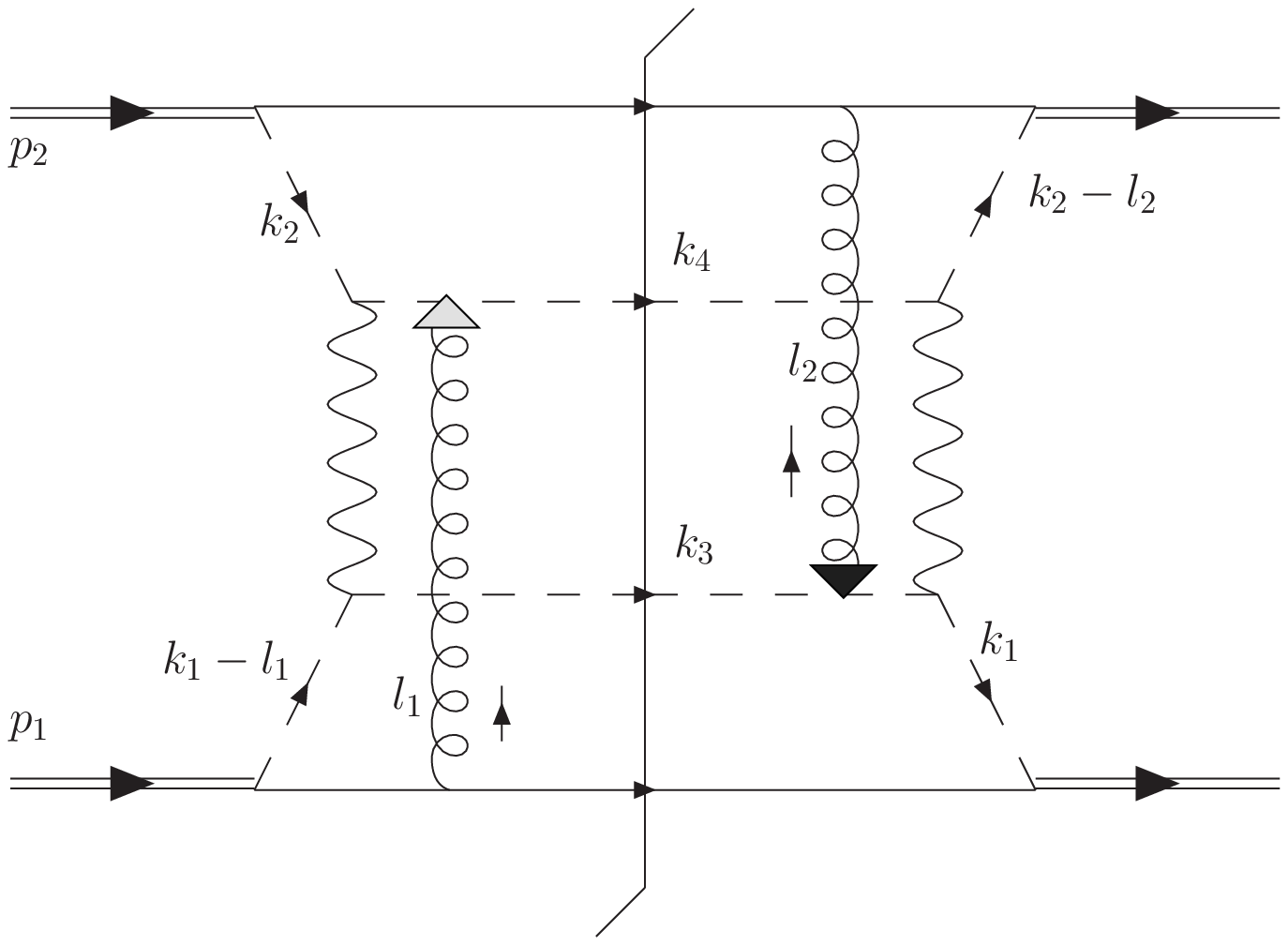}
  \\
  (a) & (b)
  \end{tabular}
\caption{Graphs that contribute to a violation of generalized TMD-factorization.    
Other graphs that should be included are those with all possible attachments of $l_1$ to the $k_4$ and $k_2$ lines, and all possible
attachments of $l_2$ to the $k_3$ and $k_1$ lines, and all Hermitian conjugate graphs.  In total there are $16$ graphs of this type.}
\label{fig:violating}
\end{figure*}

The order-$g$ contribution from a Wilson loop in Eq.~(\ref{eq:modfact}) is zero 
simply because, with one gluon attaching to the Wilson loop, the color trace in the definition of the Wilson line 
operator Eq.~(\ref{eq:loop}) involves only one color generator $\trc{t^a} = 0$.
So,
\begin{equation}
\label{eq:loopzero}
\left. U^{n_1}_{(\Box)} \right|_{\mathcal{O}(g)}  = \left. U^{n_2}_{(\Box)} \right|_{\mathcal{O}(g)} = 0.
\end{equation}
Therefore, there is no disagreement between Eq.~(\ref{eq:modfact}) and \emph{standard} factorization, Eq.~(\ref{eq:standfact}), 
from graphs with only one extra gluon.
To find a disagreement between the standard and generalized TMD-factorization formulas (Eqs.~(\ref{eq:standfact}) and~(\ref{eq:modfact})), 
at least two gluons need to be collinear
to one of the hadrons as in Sect.~\ref{sec:stanvio}.

\section{Breakdown of Generalized TMD-Factorization}
\label{sec:breakdown}

We will now directly illustrate a breakdown of the generalized TMD-factorization formula Eq.~(\ref{eq:modfact})  
by calculating the anomalous two-gluon contribution to a double Sivers effect from the graphs in Fig.~\ref{fig:violating}.  
First, let us summarize the situation so far:  
\begin{itemize}
\item 
Considering graphs with up to two extra gluons,
there is a one-to-one correspondence between graphs of the type shown in Fig.~\ref{fig:onegluestand}/Fig.~\ref{fig:twoopp} and 
the contributions to the Wilson lines
in the \emph{standard} TMD-factorization formula Eq.~(\ref{eq:standfact}).
\item The sum of graphs with two extra gluons radiated from one spectator 
and attaching on the opposite side of the 
hard subgraph (as in the graphs of Fig.~\ref{fig:twosame}) results in terms that violate standard TMD-factorization in unpolarized scattering.
These are the contributions already discussed in Ref.~\cite{Collins:2007jp}.  
\item 
The two-gluon factorization anomaly terms found in Sect.~\ref{sec:stanvio}
specify which modifications of the Wilson lines in the TMD PDFs are needed if a form of factorization is to be recovered.
It is found that the modified TMD PDFs must each contain an extra color-traced Wilson loop, as in Eq.~(\ref{eq:modTMD}).
This result also follows from the same steps for finding general Wilson line structures from low order graphs as discussed in Ref.~\cite{Bomhof:2006dp}.
\item 
Because of Eq.~(\ref{eq:loopzero}), the generalized TMD-factorizaton 
formula Eq.~(\ref{eq:modfact}) cannot include any other two-gluon contributions to Wilson lines.
In particular, there can be no contribution that corresponds to graphs like Fig.~\ref{fig:violating}, where one gluon is 
radiated collinear to each hadron simultaneously.  
These graphs would correspond to separate order-$g$ contributions from the Wilson lines in separate TMD PDFs.
If they are non-zero, then they contribute to a violation of both standard TMD-factorization in Eq.~(\ref{eq:standfact}) and the generalized 
TMD factorization formula in Eq.~(\ref{eq:modfact}).  To incorporate such a contribution, one would have to modify the Wilson line 
in each TMD PDF such that it includes a single-gluon contribution to a factorization anomaly.
But this would contradict Sect.~\ref{sec:onegluon} where it was shown that 
there is no violation of standard factorization with just one gluon.  
Hence, contributions from graphs like Fig.~\ref{fig:violating} cannot 
be consistently incorporated into a generalization of factorization simply by modifying Wilson lines in separate correlation functions.
If they give a non-vanishing contribution, then there is a clear violation of generalized TMD-factorization.
\end{itemize}
 
We will therefore prove that generalized TMD-factorization, Eq.~(\ref{eq:modfact}), is violated by showing 
that the sum of graphs of the type illustrated in Fig.~\ref{fig:violating} 
give a non-vanishing contribution to a DSA.

First, we note that all graphs of the type shown in Fig.~\ref{fig:violating} include the non-zero color factor
\begin{equation}
\label{eq:colorfact}
\trc{t^a t^b} \trc{t^b t^a} = T_F^2 (N_c^2 - 1).
\end{equation}
Next, we must ensure that there is no cancellation between graphs.

\subsection{Same Side of the Cut}

In the sum of graphs like Fig.~\ref{fig:violating}(a), where both gluons are on the same side of the cut, 
the eikonal factors give a total contribution equal to
\begin{multline}
\label{eq:same}
\left( \frac{1}{-l_1^+ + i \epsilon} \right) \left( \frac{1}{- l_2^- + i \epsilon} \right) 
+ \left( \frac{1}{l_1^+ + i \epsilon} \right) \left( \frac{1}{- l_2^- + i \epsilon} \right) + \\  +
\left( \frac{1}{-l_1^+ + i \epsilon} \right) \left( \frac{1}{l_2^- + i \epsilon} \right) 
+ \left( \frac{1}{l_1^+ + i \epsilon} \right) \left( \frac{1}{l_2^- + i \epsilon} \right) \\
= - 4 \pi^2 \delta(l_1^+) \delta(l_2^-).
\end{multline}
Since spin dependence is needed in both $H_1$ and $H_2$ for a DSA, 
then there are also two factors corresponding to Eq.~(\ref{eq:spin}) 
(but with one corresponding to a $p_2$ spectator attachment).   
Taking into account both of the resulting factors of $i$ gives an overall factor of $i^2 = -1$.  
Combined with Eq.~(\ref{eq:same}), the relevant factor from extra collinear gluons is 
then $4 \pi^2 \delta(l_1^+) \delta(l_2^-)$.  
The same result is obtained from the Hermitian conjugate graphs.

\subsection{Opposite Side of Cut}

The sum of graphs with one extra gluon on each side of the cut, as in Fig.~\ref{fig:violating}(b), works in much the same way.
The eikonal factors give
\begin{multline}
\label{eq:opp}
\left( \frac{1}{-l_1^+ - i \epsilon} \right) \left( \frac{1}{- l_2^- + i \epsilon} \right) 
+ \left( \frac{1}{l_1^+ - i \epsilon} \right) \left( \frac{1}{- l_2^- + i \epsilon} \right) + \\  +
\left( \frac{1}{-l_1^+ - i \epsilon} \right) \left( \frac{1}{l_2^- + i \epsilon} \right) 
+ \left( \frac{1}{l_1^+ - i \epsilon} \right) \left( \frac{1}{l_2^- + i \epsilon} \right) \\
= 4 \pi^2 \delta(l_1^+) \delta(l_2^-).
\end{multline}
For a DSA, there is a factor of $i$ from a factor analogous to Eq.~(\ref{eq:spin}) 
for the $p_1$-spectator attachment of the gluon on the 
left side of the cut and a factor of $-i$ from a factor analogous to Eq.~(\ref{eq:spin2}) (but for a gluon attaching 
at the $p_2$-spectator) on the right side of the cut, giving an overall factor of $i(-i) = +1$.  
So, combined with Eq.~(\ref{eq:opp}) the
relevant factor from extra gluon attachments is again $4 \pi^2 \delta(l_1^+) \delta(l_2^-)$.  

\subsection{Together}

Summing all graphs of the 
type shown in Fig.~\ref{fig:violating}, therefore, results in just a single integral.
To check it explicitly, one can use Eqs.~(\ref{eq:opp}) (extracting the overall factor of $i(-i) = 1$ that comes with 
the two spectator attachments) to explicitly calculate the contribution from graphs with gluons on opposite sides of the cut:
\begin{widetext}
\begin{multline}
\label{eq:breakingterm1}
\frac{2 T_F^2 (N_c^2 - 1) g^4}{2 s} \int \frac{d^4 k_1}{(2 \pi)^4} 
\int \frac{d^4 k_2}{(2 \pi)^4} \int \frac{d^4 l_1}{(2 \pi)^4} \int \frac{d^4 l_2}{(2 \pi)^4}
(2 \pi)^4 \delta^4 (k_1 + k_2 - k_3 - k_4) (2 \pi)^2 \delta(l_1^+) \delta(l_2^-) \times \\ \times
\left\{ {\lambda_1^\prime}^2 {\lambda_2^\prime}^2 \frac{ (k_1 + k_3) \cdot (k_2 + k_4) }{(k_1 - k_3)^2 - M_X^2} \right\}^2 
\frac{2 \epsilon_{jk} s_1^j l_1^k p_1^+ (m_{H_1} (1 - x_1) + m_{\psi_1})}{[l_1^2 + i \epsilon ] [(k_1 - l_1)^2 - m_{q_1}^2 + i \epsilon] 
[(p_1 - k_1 + l_1)^2 - m_{\psi_1}^2 + i \epsilon] [k_2^2 - m_{q_2}^2 + i \epsilon]} \times \\ \times
\frac{2 \epsilon_{j^\prime k^\prime} s_2^{j^\prime} l_2^{k^\prime} p_2^- (m_{H_2} (1 - x_2) + m_{\psi_2})}{[l_2^2 - i \epsilon ] 
[(k_2 - l_2)^2 - m_{q_2}^2 - i \epsilon] [(p_2 - k_2 + l_2)^2 - m_{\psi_2}^2 - i \epsilon] [k_1^2 - m_{q_1}^2 - i \epsilon]} (2 \pi)^2
\delta ((p_1 - k_1)^2 - m_{\psi_1}^2) \delta ((p_2 - k_2)^2 - m_{\psi_2}^2). 
\end{multline}
Here we have dropped the irrelevant factors of $(2 E_{3(4)})$, $\lambda_{1(2)}$, and $(2 \pi)^6$ which appeared
in Eq.~(\ref{eq:basiceq}).  We have also dropped the $l_1$ and $l_2$ dependence inside the factor in braces.
This is permitted because the $\delta$-functions will set $l_1^+$ and $l_2^-$ to zero, 
and because the other components yield power suppressed corrections.  
The $\delta$-functions can be used to evaluate the $k_1^+$,$k_1^-$,$k_2^+$, $k_2^-$, $l_1^+$, $l_2^-$ and ${\bf k}_{2T}$  
integrals, and the $l_1^-$ and $l_2^+$ integrals can be evaluated by contour integration.  The result is
\begin{multline}
\label{eq:breakingterm2}
\frac{T_F^2 (N_c^2 - 1) g^4}{s} \int \frac{d^2 {\bf k}_{1T}}{(2 \pi)^2} 
\int \frac{d^2 {\bf l}_{1T}}{(2 \pi)^2} \int \frac{d^2 {\bf l}_{2T}}{(2 \pi)^2} \, \mathcal{H}(k_1,k_3,k_4) \times
\\ \times
\frac{[\epsilon_{jk} s_1^j l_1^k (m_{H_1} (1 - x_1) + m_{\psi_1})]  
[\epsilon_{j^\prime k^\prime} s_2^{j^\prime} l_2^{k^\prime} (m_{H_2} (1 - x_2) + m_{\psi_2})]}{l_{1T}^2 l_{2T}^2 [k_{1T}^2 + \cdots] 
[({\bf l}_{1T} - {\bf k}_{1T})^2 + \cdots][({\bf k}_{3T} + {\bf k}_{4T} - {\bf k}_{1T})^2 + \cdots] 
[({\bf l}_{2T} - {\bf k}_{3T} - {\bf k}_{4T} + {\bf k}_{1T})^2 + \cdots]}.
\end{multline}
\end{widetext}
The factor $\mathcal{H}(k_1,k_3,k_4)$ is now the hard part in Eq.~(\ref{eq:hard}).
The symbol ``$\cdots$'' in the denominators refers to terms that 
involve only $x_1$, $x_2$ and masses. 
For graphs with both gluons on the same side of the cut, exactly the same result (with the same overall sign) is obtained
by using Eq.~(\ref{eq:same}) and taking into account the overall factor of $i^2 = -1$ that comes from the two spectator attachments.
There are, again, identical results coming from the Hermitian conjugate graphs.

The integral in Eq.~(\ref{eq:breakingterm2}) is not generally 
zero as can be checked by considering a fixed ${\bf q} \equiv {\bf k}_{3T} + {\bf k}_{4T}$.
However, there is no corresponding factorization anomaly term contained in an expansion 
of Eq.~(\ref{eq:modfact}) up to order $g^2$ in the Wilson lines.
The modified Wilson lines in the TMD PDFs of the generalized TMD-factorization 
formula are specified by the factorization anomaly terms that arise when gluons are radiated  
from just one hadron at a time, as in Sect.~\ref{sec:twogluons}.  However, the resulting generalized TMD-factorization formula is in contradiction 
with graphs like Fig.~\ref{fig:violating}, where there are simultaneously gluons radiated from $H_1$ and $H_2$.
Thus, generalized TMD-factorization breaks down at the two-gluon level.
  
Indeed, at up to two gluons, the generalized TMD-factorization formula Eq.~(\ref{eq:modfact}) gives no contribution to a 
DSA beyond 
what is predicted by the standard TMD-factorization 
formula Eq.~(\ref{eq:standfact}).
The only factorization anomaly contribution from Eq.~(\ref{eq:modfact}) is to the unpolarized 
cross section, and arises from graphs with 
both extra gluons radiated from the same hadron.  There must 
be at least one gluon coming from each 
hadron at the same time to have spin dependence in both hadrons simultaneously.
So, the only contribution to a double Sivers effect that violates standard TMD-factorization is from the sum of 
graphs in Fig.~\ref{fig:violating}, and these are 
not accounted for in Eq.~(\ref{eq:modfact}).
To recover the factorization anomaly in the DSA, one would have to somehow allow each TMD PDF to have a non-vanishing 
factorization anomaly contribution at the level of one extra gluon.  This, however, would contradict the single-extra-gluon 
treatment of the SSA in Sect.~\ref{sec:onegluon}.  Therefore, a TMD-factorization formula cannot be recovered by 
simply modifying the Wilson lines in each TMD PDF separately.  We have thus illustrated a specific example of a 
violation of generalized TMD-factorization.

\begin{figure*}
\centering
\includegraphics[scale=.7]{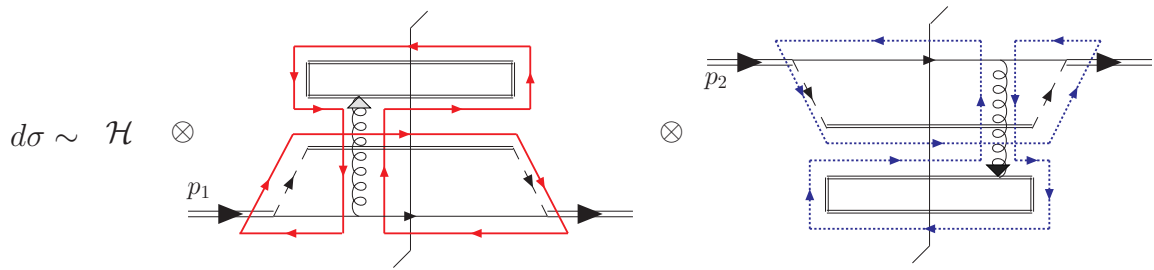}
\caption{Color flow resulting from the single gluon contributions for each of the Wilson loops 
in the TMD-factorization formula Eq.~(\ref{eq:modfact}).  
$\mathcal{H}$ is the standard zeroth order hard part and the second two factors are the TMD PDFs.
The narrow double lines represent Wilson lines.
The boxes associated with each of the TMD PDFs correspond to the Wilson loops.
The thick solid red and dotted blue lines (color online) illustrate the flow of color in each TMD PDF.
Each of the contributions to a TMD PDF shown here is exactly zero because each includes a factor $\trc{t^a} = 0$.}
\label{fig:color1}
\end{figure*}
\begin{figure*}
\centering
\includegraphics[scale=.5]{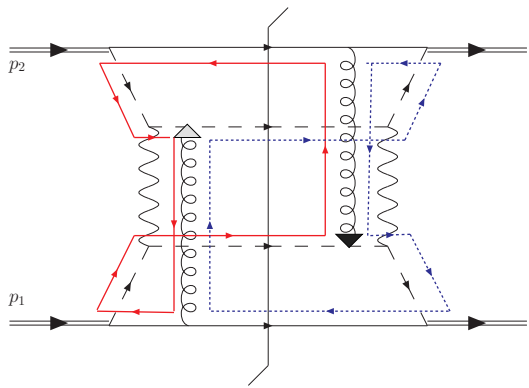}
\caption{Color flow in the unfactorized graph with a single gluon 
collinear to each of the incoming hadrons as in Fig.~\ref{fig:violating}. 
The thick solid red and dotted blue lines (color online) again illustrate the flow of color.
Non-singlet 
color can easily be exchanged and results in a non-zero contribution.  Compare with Fig.~\ref{fig:color1}.}
\label{fig:color2}
\end{figure*}
\begin{figure*}
\centering
\includegraphics[scale=.4]{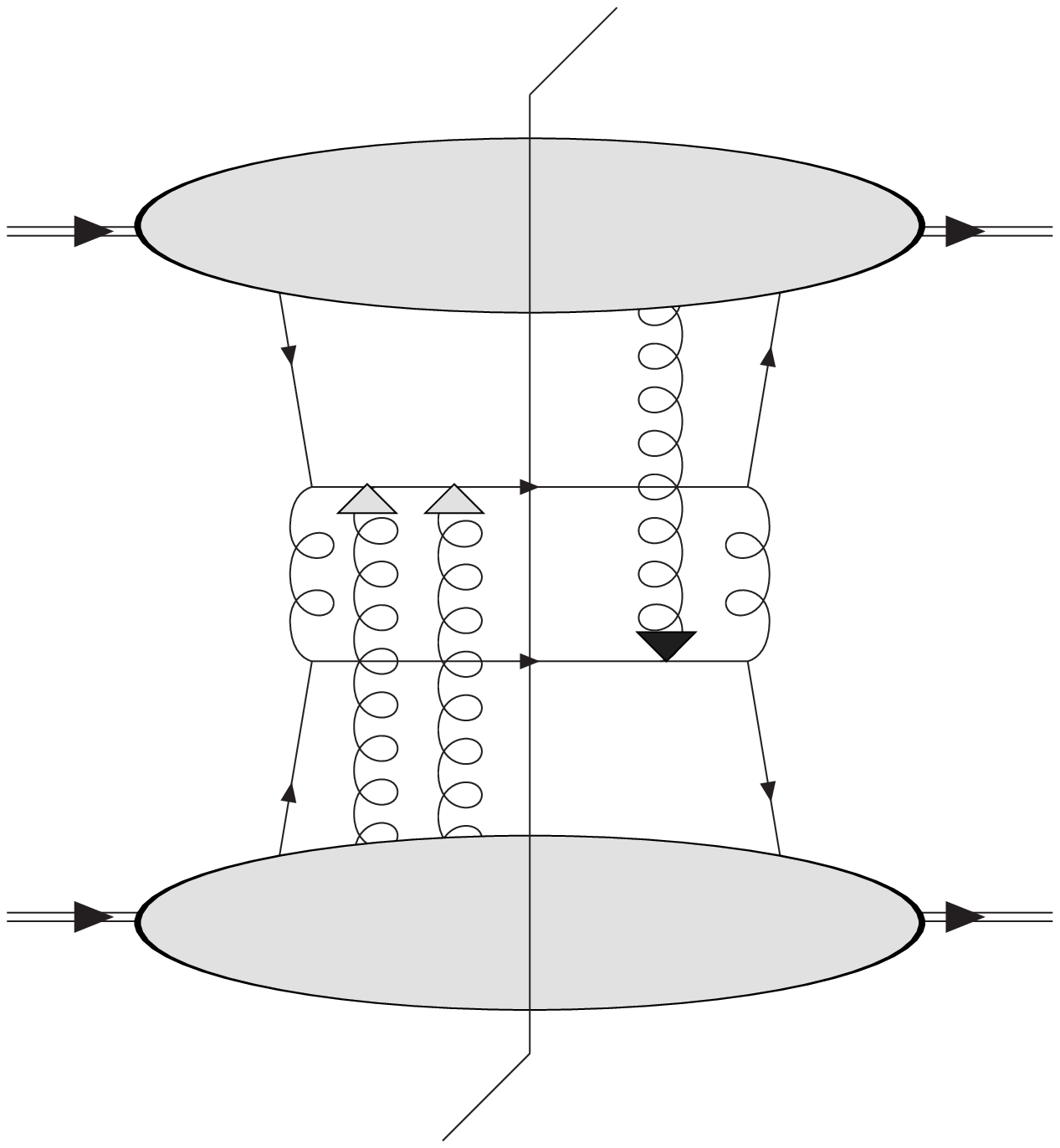}
\caption{A graph of the type which can lead to violations of generalized TMD-factorization in unpolarized scattering in real QCD.}
\label{fig:unpolgen}
\end{figure*}

One can interpret the breakdown of generalized TMD-factorization visually by imagining the flow of color in the 
factorized expression, Eq.~(\ref{eq:modfact}), as compared to the original unfactorized graph.
The graph in Fig.~\ref{fig:violating}b, for example, must correspond to the product of two single-gluon contributions 
to the Wilson lines in Eq.~(\ref{eq:modfact}), if that formula is correct.  The resulting color structure can be visualized as 
in Fig.~\ref{fig:color1}.  
The narrow double lines represent the Wilson lines in each TMD PDF.  The box shaped Wilson lines represent the extra color-traced Wilson loops in 
Eq.~(\ref{eq:modfact}).  The color trace over each of the Wilson loops means that only a color singlet 
gluon can be exchanged between a Wilson loop and a spectator 
at the single-gluon level, as illustrated by the factors representing 
the TMD PDFs in Fig.~\ref{fig:color1}.  Therefore, these contributions to the TMD PDFs must vanish.
By contrast, in the original unfactorized graph, the interlaced color flow of the two 
gluons means there is no problem with exchanging non-singlet color, as illustrated by 
the color flow diagram in Fig.~\ref{fig:color2}.  This corresponds to the non-zero 
color factor in Eq.~(\ref{eq:colorfact}).

\section{Discussion and Conclusion}
\label{sec:conc}

We have shown that a generalized TMD-factorization formula like Eq.~(\ref{eq:genfactor})/Eq.~(\ref{eq:modfact}) does not in general exist 
for back-to-back hadro-production of high-$p_t$ hadrons due to the non-trivial interplay of gluons that are collinear 
to both hadron directions.  The failure of generalized TMD-factorization calls into question results that use it as a starting assumption. 
This includes calculations of weighted cross sections which retain a memory of the non-universality of the TMD PDFs. 
 
It is interesting to note that the same complications do not seem to arise if the gauge field is Abelian.  In that case, the 
Abelian ``color factors'' can be identified with coupling constants, such as in the model from Ref.~\cite{collins.qiu}. 
Then the contribution 
to a DSA from graphs like Fig.~\ref{fig:violating} is indeed just a product of single gluon contributions to Wilson loops from separate
correlation functions (which are non-zero in the Abelian case).  Therefore, there is no obvious
contradiction with generalized TMD-factorization for the Abelian case, at least at the level of graphs discussed here. 
While a full proof of generalized TMD-factorization for the Abelian gauge theory does not yet exist,
the breakdown of generalized TMD-factorization that we have found is a specific consequence of the non-Abelian nature of QCD.

It is worth stressing that the failure of generalized TMD-factorization occurs in a regime where factorization would 
ordinarily be expected to apply; namely for hard processes with large-$p_t$.
Moreover, it should not be thought that the breakdown of generalized TMD-factorization is specific to a DSA or SSA.  We have 
calculated for a DSA  with color singlet boson exchange for the hard part in order to maximize the simplicity of the argument, but the reasons for the 
breakdown are quite general.  
The counter-example we have provided in this paper is sufficient to show that a general proof does not exist.  That is, 
it is not possible to consistently define TMD PDFs for each hadron separately, even if we allow for process dependence 
in the Wilson line structures. 
To encounter the same complications with generalized TMD-factorization in an 
unpolarized cross section, one needs the type of non-cancellation that was found in Sect.~\ref{sec:stanvio} (which comes from 
having two gluons collinear to one hadron), but with at least one more gluon 
collinear to the opposite hadron.  An example of the type of graph which can produce a violation of generalized TMD-factorization 
in the unpolarized cross section is shown in Fig.~\ref{fig:unpolgen}.

Furthermore, the contribution from extra gluon attachments (which lead to factorization breaking) should not be 
thought of as negligible higher order corrections.  They
correspond to soft and collinear divergences in higher order 
hard scattering calculations.  In real QCD, they are non-perturbative gluons in the 
strong coupling regime.  
For the perturbation series 
to have sensible convergence properties, it must be possible to rearrange terms such that the extra soft and collinear gluons 
are resummed to all orders into 
TMD PDFs.
We also remark that, at the level of many extra 
gluons, there are more graphs with gluons collinear to both hadrons simultaneously (which led to a breakdown of generalized 
TMD-factorization) than graphs with gluons radiated from one hadron only.  

While the observation of generalized TMD factorization breaking leads to
frustrating practical difficulties in cross section calculations,
it should not necessarily be regarded as a purely negative result. 
The question of whether transverse momentum effects can be meaningfully associated
with parton transverse momentum in separate parton correlation functions for each hadron is intrinsically important in the search 
for an improved fundamental understanding of QCD dynamics in hard collisions. 
A counter-proof implies the existence of effects which challenge normal partonic intuition, 
and suggests new avenues of research.  While naive factorization fails, 
the fact that the extra gluons eikonalize suggests that Wilson lines still may 
play a role.  Insight might be gained, for example, from methods currently 
being applied to small-$x$ physics (e.g.~\cite{Gelis:2008rw,Balitsky:2001gj}). 
Another possibility is to model factorization breaking effects 
by directly calculating factorization breaking phase contributions in perturbation 
theory, but with explicit infrared and collinear cutoffs.
   
It is possible to understand the origin of the generalized TMD-factorization breakdown intuitively 
as arising from non-linear effects in the phases acquired by partons as they pass through the $A^+$ and $A^-$ fields of the colliding hadrons.   
If the overall phase were simply the product  
of the phases induced by the $A^+$ fields from hadron $H_1$ and the $A^-$ fields from 
hadron $H_2$, 
then one could associate any process-dependent phases induced by the $A^+$ field in hadron $H_1$ with 
a modified Wilson line for the TMD PDF of $H_1$ and, likewise, any process-dependent phases 
induced by the $A^-$ field from $H_2$ could be associated with a modified Wilson line for the TMD PDF of $H_2$. 
However, in the non-Abelian theory the role of the $A^-$ gluons in $H_2$ 
is affected by the presence of the $A^+$ gluons from $H_1$ 
and visa-versa.  A direct example of this is Fig.~\ref{fig:violating}/Eq.~(\ref{eq:breakingterm2}), 
where a single $A^-$ gluon exchanged between $H_2$ and 
the opposite-side struck quark gives a non-zero contribution, but only 
because there is simultaneously an $A^+$ gluon exchanged between $H_1$ and 
the other struck quark.
This means that one cannot address the role of phases induced by the $A^+$ and $A^-$ fields independently, but instead must deal with 
them simultaneously.
The result is a kind of nonperturbative correlation which cannot be identified as 
arising strictly from gluons coming from either hadron independently, but only from the combination.

\section*{Acknowledgments}

We would especially like to thank D.~Boer and C.~Mantz for useful discussions.
One of us (T.~Rogers) would like to thank J.~Collins for helpful discussions.
Support was provided by the research program of the ``Stichting voor Fundamenteel Onderzoek der Materie (FOM)'', 
which is financially supported by the ``Nederlandse Organisatie voor Wetenschappelijk Onderzoek (NWO)''.
All figures were made using Jaxodraw~\cite{jaxo}.

\end{document}